\documentclass[10pt]{article}

\usepackage{graphicx}
\usepackage{authblk}
\usepackage{natbib}
\usepackage{amssymb,amsmath}
\usepackage{hyperref}
\usepackage{longtable}
\usepackage{lscape}
\usepackage{url}
\usepackage[margin=1.4cm]{geometry}
\usepackage{abstract}

\title{The Influential Roles of Gravity, Turbulence, and Magnetic Fields in Shaping the Physical Evolution of Dense Massive Clumps}

\author[1]{Moses Onyemaechi Asogwa}
\author[2]{Seblu Humne Negu}
\author[2]{Gemechu Muleta Kumssa}
\author[3]{Innocent Okwudili Eya}

\affil[1]{Department of Physics, College of Natural and Computational Sciences, Addis Ababa University, Ethiopia. Email: moses.asogwa@aau.edu.et}
\affil[2]{Astronomy and Astrophysics Research Division, Entoto Observatory and Research Center (EORC), Ethiopian Space Science and Technology Institute (ESSTI), Addis Ababa, Ethiopia}
\affil[3]{Department of Physics and Astronomy, University of Nigeria, Nsukka, Nigeria}

\date{}

\begin{document}
\maketitle

\begin{abstract}
\noindent We explore the roles of the three competitors, namely, gravity, turbulence, and magnetic fields, in controlling star formation (SF) within dense, massive clumps identified in the ATLASGAL survey. By examining scaling relations, virial parameters, and turbulent energy spectra, we evaluate the dynamical state of these clumps. We observe a weak velocity dispersion-size relation ($\sigma \propto L^{0.11}$), which is much shallower than the classical Larson-like relations, suggesting that turbulence does not mainly drive internal dynamics. The turbulent energy spectrum, $E(k) \propto k^{-1.21}$, is also less steep than what is expected for both incompressible and compressible turbulence. We equally observe a decreasing trend in the virial parameter with increasing mass ($\alpha_{\mathrm{vir}} \propto M^{-0.37}$), indicating that more massive clumps are increasingly gravitationally bound. These trends indicate an increasing relative dominance of gravity over turbulence at smaller scales, aligning with multiscale collapse scenarios; however, the absolute energy balance remains unquantifiable with the current data. Although magnetic fields are not directly measured, their potential influence is considered in the interpretation of pressure balance and dynamical support. Our findings imply that gravitational processes appear to primarily regulate the structure and evolution of massive clumps.
\end{abstract}

\begin{quote}
\noindent\textbf{Key words:} stars: formation - stars: kinematics and dynamics - stars: massive - ISM: clouds - ISM: kinematics and dynamics - turbulence
\end{quote}

\section{Introduction}           
\label{sect:intro}

\noindent Stars emerge from molecular clouds (MCs), which are intricate systems denoted by stratified density structures like filaments, clumps, and cores (e.g., \citealt{Luo+etal+2024a}). Understanding these structures is essential, as they create the necessary conditions for SF. Clumps and cores are recognised as the primary sources of diverse stellar configurations (\citealt{Williams+etal+1994, Phelps+Lada+1997}), encompassing clusters, small groups, and isolated stars within the interstellar medium (ISM) (\citealt{Gammie+etal+2003}).

The mechanisms underlying low-mass SF are well understood; however, high-mass SF requires further exploration (\citealt{Motte+etal+2018}). Recent studies indicate that high-mass SF occurs across various scales, engaging stratified density structures in a top-down manner, from filaments to the initial stages of SF (e.g., \citealt{Peretto+etal+2013, Motte+etal+2018, Vazquez-Semadeni+etal+2019, Liu+etal+2022a,  Liu+etal+2022b, Liu+etal+2023, Yang+etal+2023, Luo+etal+2024a}). This multiscale phenomenon is influenced by key factors such as gravity, turbulence, and magnetic fields, along with the effects of protostellar feedback and the contributions of prior stellar generations (\citealt{Liu+etal+2022mag}).

Two primary models have been proposed by researchers to elucidate the functioning of MCs and the factors influencing SF: the ``global hierarchical collapse (GHC)" and the ``inertial inflow (I2)" models (\citealt{Vazquez-Semadeni+etal+2019, Padoan+etal+2020, Luo+etal+2024a}). Both models aim to elucidate the complex interactions among turbulence, gravity, and feedback processes that regulate SF, particularly in the context of massive stars. Their perspectives vary regarding the effects of turbulence, the environmental conditions during formation, and the fundamental processes responsible for gas sinking and mass accumulation.

The GHC model  (\citealt{Vazquez-Semadeni+etal+2019}) suggests that the initial MC is in a state close to virial equilibrium, typically characterised by subsonic or transonic turbulence (\citealt{Vazquez-Semadeni+etal+2019}). It emphasises the significance of gravitational self-interactions in gas infall dynamics across different scales. The I2 model posits that turbulence is more prevalent at larger scales (\citealt{Luo+etal+2024a}), indicating that the initial MC is significantly turbulent and not in a state of equilibrium (\citealt{Padoan+etal+2020}). This model interprets turbulence as supersonic, attributing large-scale gas infall to the inertial influx of turbulent gas.

Magnetic fields significantly influence these processes by preventing disintegration due to gravity (\citealt{Shu+etal+1987}) and inducing non-uniform turbulence (\citealt{Goldreich+Sridhar+1995}). Observational investigations of magnetic fields are essential for a more profound understanding, as they will aid in differentiating between strong-field and weak-field theories of SF (\citealt{Luo+etal+2024a}). Strong-field theories suggest that magnetically subcritical clouds develop supercritical substructures that ultimately undergo collapse (\citealt{Mouschovias+etal+2006}), while weak-field theories contend that extensive supersonic turbulent flows create overdense regions that collapse dynamically (\citealt{MacLow+Klessen+2004}).
\begin{equation}
    \sigma \propto L^{\alpha}
\end{equation}
and
\begin{equation}
    \rho \propto L^{-\beta},
\end{equation}
where the scaling exponents are denoted by $\alpha$ and $\beta$, and $L$ is the size of the structures that produce the density of clouds. Eqs. 1 and 2 represent two of the three widely recognized \cite{Larson+1981} scaling relations.

Larson relations (\citealt{Larson+1981}) have been extensively studied in star-forming clouds (e.g., \citealt{Solomon+etal+1987, Shetty+etal+2012}), especially high-mass stars ($M_{\mathrm{\star}}\, > \, 8 M_{\odot}$) (\citealt{Caselli+Myers+1995, Traficante+etal+2018a, Li+etal+2023}). The velocity dispersion-size relationship, as described in Eq. 1, aligns with the turbulence theory, which posits that energy is introduced at larger scales and subsequently cascades to smaller scales, resulting in dynamic eddies that energise the cloud (e.g., \citealt{Schneider+etal+2011}). The exponent $\alpha$ varies among different turbulence types, indicating a complex relationship among factors such as sensitivity and selection (\citealt{Larson+1981, Kegel+1989}), cloud stability (\citealt{Chieze+1987}), magnetic and kinetic energy equipartition (\citealt{Myers+Goodman+1988a}), external pressure (\citealt{Fleck+1988, Elmegreen+1989}), and the critical influence of SF (\citealt{McKee+1989}). \cite{Larson+1981} identified scaling exponents $\alpha$ approximately equal to 0.38 and $\beta$ approximately equal to 1.10, which are analogous to incompressible turbulence. The observed $\alpha$ value is close to the theoretical value of $\tfrac{1}{3}$ proposed by \cite{Kolmogorov+1941} (\citealt{Feng+etal+2024}). However, interstellar gas may demonstrate compressibility, with $\alpha$ expectations exceeding $\tfrac{1}{3}$ (\citealt{Cen+2021}), underscoring the complexity of the universe.

The value of $\alpha$ for MCs in the Milky Way (MW) disc has been measured to be approximately 0.63 (\citealt{Miville-Deschenes+etal+2017}). Other reported values of $\alpha$ are 0.21 (\citealt{Caselli+Myers+1995}), 0.56 (\citealt{Heyer+Brunt+2004}), 0.70 (\citealt{Zhou+etal+2022}), and 0.30 (\citealt{Barman+etal+2025}). These measures, obtained from various tracers across multiple objects, enrich the discussion while complicating direct comparisons. The consensus indicates that the $\alpha$ value should approximate 0.50 (\citealt{Solomon+etal+1987}), consistent with the principles governing compressible turbulence. The wide range of $\alpha$ and $\beta$ values highlights the complex impact of cloud gravity, especially in high-mass star-forming MCs (\citealt{Peretto+etal+2023}). Recent studies conducted by \cite{Barman+etal+2025} used $^{12}$CO to explore small MCs and C$^{18}$O for cores. The MCs exhibited $\alpha = 0.30 \pm 0.04$, indicating that turbulence significantly influences the dynamics of cloud structure (\citealt{Barman+etal+2025}). No apparent trend or connection was identified for the cores, suggesting that dense core areas may have a more complex link between turbulence and size (\citealt{Barman+etal+2025}). The authors found an anti-correlation ($\beta = -1.34 \pm 0.14$) between the size and density of MCs, which means that the surface density is almost always the same. For the cores, $\beta = -2.02 \pm 0.04$, which deviates significantly from a constant surface density (\citealt{Barman+etal+2025}). These findings show a multi-scale mechanism, including gravity, turbulence, and perhaps magnetic fields, affects MC SF (e.g.,\citealt{Liu+etal+2022mag}; \citealt{Luo+etal+2024a}; \citealt{Barman+etal+2025}). The ongoing discussion regarding the relative importance of these forces in MCs informs future research directions.

This study initiates an investigation into the Larson-type correlations between velocity dispersion and density in a sample of 179 dense massive clumps in the MW, as identified by \cite{Heyer+etal+2016}. This study explores the roles of the three competitors, namely, gravity, turbulence, and magnetic fields, in regulating SF within these clumps across various evolutionary phases. We designed this article's framework to facilitate your engagement with this study. Section 2 presents the data set. Section 3 examines the theoretical framework underlying the physical characteristics of the clumps and the general Larson-type relations.  Section 4 provides a detailed analysis of the observed relations, focusing on the energy spectrum of turbulence, assessment of the role of magnetic fields in maintaining clump stability and inferences on the formation of high-mass stars. Finally, Section 5 summarizes the findings and concludes the study.

\section{Data Sets}
\label{sect:dis}

\noindent The investigation into the origins of high-mass stars has prompted a series of innovative efforts focused on accurately mapping the Galactic plane at millimetre and submillimeter wavelengths (\citealt{Contreras+etal+2013}). The APEX Telescope Large Area Survey of the Galaxy (ATLASGAL; \citealt{Schuller+etal+2009}) represents the inaugural comprehensive survey elucidating the dynamics of the Galactic plane, providing valuable insights into the cores and clusters responsible for SF.

This study explores new avenues through the application of the Large APEX Bolometer Camera (LABOCA; \citealt{Siringo+etal+2009}), which features 295 bolometers operating at 870 $\mu$m (345 GHz). Sub-mm surveys are capable of detecting warm dust across a broad spectrum of temperatures, including the coldest dust. This differs from shorter wavelength surveys such as the Galactic Legacy Infrared Mid-Plane Survey Extraordinaire (GLIMPSE; \citealt{Benjamin+etal+2003}) and the MIPS Inner Galactic Plane Survey (MIPSGAL; \citealt{Carey+etal+2009}). This illustrates the initial phases of SF (\citealt{Contreras+etal+2013}).

The dataset is detailed in Table 1. The study encompasses massive, dense clumps at various evolutionary stages, exhibiting significant variation in density structures ranging from 2.50 $\mathrm{pc}$ to 0.05 $\mathrm{pc}$. This variant facilitates the examination of the connections between $\sigma$-$L$ and $\rho$-$L$, which are associated with the various evolutionary stages of these clumps.

We carefully extracted critical data, including the mass ($M_{\mathrm{cl}}$) and kinematic distance ($D$) of 179 ATLASGAL dense massive clumps that \cite{Heyer+etal+2016} initially investigated. We were able to obtain a comprehensive understanding of these stellar nurseries by utilising the data obtained from high-density molecules such as $\mathrm{NH_3}$ (\citealt{Dunham+etal+2011, Wienen+etal+2012,  Wienen+etal+2018}), $\mathrm{HCO}^+$ (\citealt{Shirley+etal+2013}), $\mathrm{N_2H}^+$ (\citealt{Jackson+etal+2013, Shirley+etal+2013, Urquhart+etal+2019}), HCN (\citealt{Jackson+etal+2013}), and CS (\citealt{Jackson+etal+2008, Urquhart+etal+2014a}). Additionally, we meticulously extracted additional data, including the semi-major ($a$) and minor ($b$) axes (\citealt{Urquhart+etal+2014a}), dust temperature ($T_{\mathrm{dust}}$) (\citealt{Contreras+etal+2013}), and kinematic temperature ($T_{\mathrm{kin}}$) (\citealt{Wienen+etal+2012}). As we investigate the extraordinary realm of SF, these critical parameters assist us in comprehending the physical characteristics of these stunning clumps.

Massive clumps represent places of maximal turbulence; hence, their velocity dispersion measurements do not encompass the whole turbulence, but turbulence does impact their behaviour (\citealt{Lu+etal+2022}). Utilizing appropriate molecular line tracers (e.g., NH$_{3}$, HCO$^{+}$, N$_{2}$H$^{+}$, HCN, and CS) for assessing the turbulence features of dense massive clumps is a rational approach to investigate varying density conditions within the area. Using information from these molecular lines to examine the scaling relations and the corresponding turbulent energy spectrum is essential, as this methodology pertains to the physically relevant gas regimes defined by the interplay of turbulence and gravity, consistent with established observational and theoretical frameworks (e.g., \citealt{Goodman+etal+1998}; \citealt{Wu+etal+2010}; \citealt{Gaches+etal+2015}; \citealt{Orkisz+etal+2017}; \citealt{Yun+etal+2021}; \citealt{Luo+etal+2024a}; \citealt{Yang+etal+2025}).

\newcommand{\tableName}{}
\begin{landscape}
\scriptsize
\begin{longtable}{|c|c|c|c|c|c|c|c|c|c|c|c|c|} 
\caption{Calculated properties of resolved ATLASGAL star-forming clumps with distance measure.} \label{tab:my_longtable} \\

\hline
ATLASGAL name & $M_{\mathrm{cl}}$ & $\Sigma_{\mathrm{H_2}}$ & $\sigma$ & $\alpha_{\mathrm{vir}}$ & $a$ & $b$ & $L{\mathrm{cl}}$ & $P_{\mathrm{turb}}$ & $P_{\mathrm{grav}}$ & $\epsilon_{\mathrm{k}}$ & Tracer & Evolutionary phase \\ [0.5ex]
 & ($M_{\odot}$) & ($M_{\odot}\,\mathrm{pc}^{-2}$) & ($\mathrm{km}\,\mathrm{s}^{-1}$) & & ($\mathrm{arcsec}$) & ($\mathrm{arcsec}$) & ($\mathrm{pc}$) & ($M_{\odot}\,\mathrm{pc}^{-1}\,\mathrm{s}^{-2}$) & ($M_{\odot}\,\mathrm{pc}^{-1}\,\mathrm{s}^{-2}$) & ($\mathrm{erg}\,\mathrm{pc}^{-3}\,\mathrm{s}^{-1}$) & & \\ [0.5ex]
\hline
\endfirsthead
\caption[]{(Continued) Calculated properties of resolved ATLASGAL star-forming clumps with distance measure.} \\
\hline
ATLASGAL name & $M_{\mathrm{cl}}$ & $\Sigma_{\mathrm{H_2}}$ & $\sigma$ & $\alpha_{\mathrm{vir}}$ & $a$ & $b$ & $L{\mathrm{cl}}$ & $P_{\mathrm{turb}}$ & $P_{\mathrm{grav}}$ & $\epsilon_{\mathrm{k}}$ & Tracer & Evolutionary phase \\ [0.5ex]
 & ($M_{\odot}$) & ($M_{\odot}\,\mathrm{pc}^{-2}$) & ($\mathrm{km}\,\mathrm{s}^{-1}$) & & ($\mathrm{arcsec}$) & ($\mathrm{arcsec}$) & ($\mathrm{pc}$) & ($M_{\odot}\,\mathrm{pc}^{-1}\,\mathrm{s}^{-2}$) & ($M_{\odot}\,\mathrm{pc}^{-1}\,\mathrm{s}^{-2}$) & ($\mathrm{erg}\,\mathrm{pc}^{-3}\,\mathrm{s}^{-1}$) & & \\ [0.5ex]
\hline
\endhead
\hline \hline
AGAL005.397+00.194 &	142 &	311 &	0.52 &	0.83 &	22 &	11 &	0.20 &	5.79E-26 &	2.16E-26 &	2.77E+32 &	$\mathrm{NH_3}\,(1,1)$ &	Protostellar \\
AGAL005.617-00.082 &	37030 &	429 &	0.94 &	0.14 &	30 &	14 &	2.50 &	1.89E-26 &	3.38E-26 &	1.31E+31 &	$\mathrm{NH_3}\,(1,1)$ &	Protostellar \\
AGAL007.166+00.131 &	2223 &	1274 &	0.80 &	0.25 &	11 &	9 &	0.49 &	2.87E-25 &	5.68E-25 &	8.64E+32 &	$\mathrm{NH_3}\,(1,1)$ &	Protostellar \\
AGAL007.333-00.567 &	567 &	457 &	0.49 &	0.31 &	21 &	16 &	0.30 &	4.58E-26 &	3.85E-26 &	1.38E+32 &	$\mathrm{NH_3}\,(1,1)$ &	YSO \\
AGAL007.636-00.192 &	1276 &	692 &	0.65 &	0.30 &	12 &	10 &	0.49 &	1.00E-25 &	1.58E-25 &	2.45E+32 &	$\mathrm{NH_3}\,(1,1)$ &	YSO \\
AGAL008.206+00.191 &	90 &	622 &	0.46 &	0.57 &	12 &	10 &	0.13 &	1.61E-25 &	1.15E-25 &	1.05E+33 &	$\mathrm{NH_3}\,(1,1)$ &	Protostellar \\
AGAL008.544-00.341 &	374 &	759 &	0.54 &	0.36 &	16 &	9 &	0.25 &	1.47E-25 &	1.86E-25 &	5.85E+32 &	$\mathrm{NH_3}\,(1,1)$ &	Protostellar \\
AGAL008.706-00.414 &	15356 &	486 &	0.53 &	0.07 &	39 &	16 &	1.41 &	1.13E-26 &	3.77E-26 &	7.82E+30 &	$\mathrm{NH_3}\,(1,1)$ &	Protostellar \\
AGAL008.804-00.327 &	315 &	263 &	1.04 &	2.47 &	20 &	13 &	0.31 &	1.21E-25 &	1.41E-26 &	7.46E+32 &	$\mathrm{HCO}^+\,(3-2)$ &	Protostellar \\
AGAL008.954-00.532 &	9264 &	261 &	1.30 &	0.71 &	30 &	19 &	1.49 &	3.44E-26 &	1.08E-26 &	5.53E+31 &	$\mathrm{NH_3}\,(1,1)$ &	Protostellar \\
AGAL009.284-00.147 &	683 &	401 &	0.79 &	0.79 &	33 &	11 &	0.39 &	8.92E-26 &	3.65E-26 &	3.32E+32 &	$\mathrm{N_2H}^+\,(1-0)$ &	Protostellar \\
AGAL009.851-00.031 &	104 &	626 &	0.39 &	0.40 &	25 &	9 &	0.16 &	1.09E-25 &	1.53E-25 &	4.88E+32 &	$\mathrm{NH_3}\,(1,1)$ &	YSO \\
AGAL009.966-00.021 &	2463 &	3054 &	0.74 &	0.13 &	10 &	9 &	0.53 &	8.66E-25 &	8.25E-24 &	2.23E+33 &	$\mathrm{N_2H}^+\,(1-0)$ &	YSO \\
AGAL010.404-00.201 &	309 &	640 &	0.54 &	0.43 &	22 &	16 &	0.18 &	1.25E-25 &	6.98E-26 &	6.91E+32 &	$\mathrm{NH_3}\,(1,1)$ &	Protostellar \\
AGAL010.742-00.126 &	781 &	470 &	0.53 &	0.30 &	22 &	18 &	0.34 &	4.76E-26 &	3.91E-26 &	1.37E+32 &	$\mathrm{NH_3}\,(1,1)$ &	Protostellar \\
AGAL010.991-00.082 &	702 &	300 &	0.67 &	0.63 &	38 &	19 &	0.38 &	4.10E-26 &	1.41E-26 &	1.33E+32 &	$\mathrm{NH_3}\,(1,1)$ &	Quiescent \\
AGAL011.004-00.071 &	470 &	858 &	0.84 &	0.73 &	21 &	10 &	0.24 &	3.80E-25 &	1.97E-25 &	2.45E+33 &	$\mathrm{N_2H}^+\,(1-0)$ &	YSO \\
AGAL011.064-00.099 &	498 &	611 &	1.04 &	1.30 &	31 &	11 &	0.27 &	3.41E-25 &	8.47E-26 &	2.42E+33 &	$\mathrm{HNC}\,(1-0)$ & Protostellar \\
AGAL011.082-00.534 &	579 &	682 &	0.92 &	0.88 &	20 &	15 &	0.25 &	2.92E-25 &	8.69E-26 &	1.97E+33 &	$\mathrm{HNC}\,(1-0)$ & Protostellar \\
AGAL011.597-00.131 &	2654 &	617 &	1.68 &	1.44 &	15 &	10 &	0.72 &	3.91E-25 &	1.17E-25 &	1.68E+33 &	$\mathrm{HCO}^+\,(3-2)$ &	YSO \\
AGAL012.496-00.222 &	362 &	450 &	0.61 &	0.61 &	19 &	13 &	0.25 &	8.69E-26 &	4.00E-26 &	3.91E+32 &	$\mathrm{NH_3}\,(1,1)$ &	Quiescent \\
AGAL012.554-00.347 &	220 &	767 &	1.08 &	1.85 &	14 &	10 &	0.19 &	7.77E-25 &	1.88E-25 &	8.14E+33 &	$\mathrm{HCO}^+\,(3-2)$ &	Protostellar \\
AGAL012.914-00.336 &	208 &	507 &	1.13 &	2.54 &	23 &	10 &	0.20 &	4.70E-25 &	6.36E-26 &	4.88E+33 &	$\mathrm{N_2H}^+\,(1-0)$ &	Protostellar \\
AGAL012.954-00.229 &	786 &	192 &	1.97 &	6.53 &	41 &	27 &	0.49 &	1.71E-25 &	5.49E-27 &	1.27E+33 &	$\mathrm{HCO}^+\,(3-2)$ &	Protostellar \\
AGAL012.988+00.354 &	77 &	1452 &	1.00 &	1.95 &	14 &	8 &	0.14 &	2.93E-24 &	1.99E-24 &	3.85E+34 &	$\mathrm{HCO}^+\,(3-2)$ &	Protostellar \\
AGAL013.038-00.314 &	109 &	686 &	0.96 &	2.16 &	15 &	9 &	0.15 &	7.39E-25 &	1.69E-25 &	8.70E+33 &	$\mathrm{N_2H}^+\,(3-2)$ &	Protostellar \\
AGAL013.109-00.217 &	183 &	436 &	0.78 &	1.43 &	12 &	10 &	0.24 &	1.91E-25 &	6.63E-26 &	1.14E+33 &	$\mathrm{N_2H}^+\,(3-2)$ &	Protostellar \\
AGAL013.259-00.409 &	579 &	459 &	0.87 &	0.96 &	18 &	17 &	0.29 &	1.44E-25 &	3.57E-26 &	7.96E+32 &	$\mathrm{NH_3}\,(1,1)$ &	HII region \\
AGAL013.349+00.196 &	67 &	568 &	0.75 &	1.86 &	15 &	10 &	0.11 &	4.33E-25 &	8.41E-26 &	5.44E+33 &	$\mathrm{HCO}^+\,(3-2)$ &	Protostellar \\
AGAL014.282-00.511 &	82 &	786 &	0.41 &	0.42 &	23 &	12 &	0.09 &	1.91E-25 &	1.22E-25 &	1.60E+33 &	$\mathrm{N_2H}^+\,(3-2)$ &	YSO \\
AGAL014.622-00.131 &	148 &	450 &	0.87 &	1.90 &	14 &	10 &	0.19 &	2.76E-25 &	5.65E-26 &	2.33E+33 &	$\mathrm{N_2H}^+\,(3-2)$ &	Quiescent \\
AGAL014.707-00.156 &	582 &	352 &	0.82 &	0.99 &	27 &	15 &	0.33 &	8.56E-26 &	2.08E-26 &	3.91E+32 &	$\mathrm{NH_3}\,(1,1)$ &	Protostellar \\
AGAL014.851-00.990 &	217 &	1067 &	0.60 &	0.49 &	15 &	11 &	0.14 &	3.96E-25 &	2.79E-25 &	3.12E+33 &	$\mathrm{NH_3}\,(1,1)$ &	YSO \\
AGAL014.949-00.072 &	196 &	853 &	1.37 &	3.01 &	14 &	11 &	0.16 &	1.55E-24 &	2.06E-25 &	2.45E+34 &	$\mathrm{HCO}^+\,(3-2)$ &	YSO \\
AGAL015.531-00.407 &	382 &	452 &	0.52 &	0.43 &	17 &	15 &	0.25 &	6.19E-26 &	3.83E-26 &	2.37E+32 &	$\mathrm{NH_3}\,(1,1)$ &	YSO \\
AGAL016.318-00.531 &	7069 &	939 &	0.54 &	0.08 &	19 &	11 &	0.86 &	4.64E-26 &	2.20E-25 &	5.38E+31 &	$\mathrm{NH_3}\,(1,1)$ &	YSO \\
AGAL016.418-00.634 &	1308 &	654 &	0.71 &	0.36 &	29 &	15 &	0.36 &	1.08E-25 &	7.04E-26 &	3.94E+32 &	$\mathrm{NH_3}\,(1,1)$ &	Quiescent \\
AGAL016.442-00.384 &	178 &	606 &	0.45 &	0.41 &	16 &	9 &	0.20 &	1.05E-25 &	1.27E-25 &	4.38E+32 &	$\mathrm{NH_3}\,(1,1)$ &	Protostellar \\
AGAL016.459-00.669 &	292 &	536 &	0.69 &	0.80 &	15 &	11 &	0.22 &	1.61E-25 &	6.49E-26 &	9.29E+32 &	$\mathrm{NH_3}\,(1,1)$ &	YSO \\
\hline 
\caption*{Column 1, is the ATLASGAL name, $M_{\mathrm{cl}}$ (column 2), is the clump mass, $\Sigma_{\mathrm{H_2}}$ (column 3), is the surface density, $\sigma$ (column 4), is the velocity dispersion, $\alpha_{\mathrm{vir}}$ (column 5), is the virial parameter. $a$ (column 6), is the semi-major axis, $b$ (column 7), is the semi-minor axis, $L{\mathrm{cl}}$ (column 8), is the linear size, $P_{\mathrm{turb}}$ (column 9), is the turbulent pressure, $P_{\mathrm{grav}}$ (column 10), refers to the gravitational pressure, $\epsilon_{\mathrm{k}}$ (column 11), is the turbulent kinetic energy transfer rate per unit volume, and columns 12 and 13, refer to the molecular tracer and clumps evolutionary phases, respectively. The complete table is available at  \href{https://doi.org/10.57760/sciencedb.32759}{https://doi.org/10.57760/sciencedb.32759}.}
\end{longtable}
\end{landscape}

\subsection{Clumps categorization}
\noindent The clumps five distinct evolutionary phases categorization were extracted from the ATLASGAL Compact Source Catalogue (CSC; \citealt{Contreras+etal+2013, Urquhart+etal+2014b, Urquhart+etal+2022}), identified using 8, 24, and 70 $\mu$m wavelengths.

\begin{enumerate}
	\item Quiescents: Cold clumps (10-15 K) without any embedded objects and dark at 70 			$\mu$m.
	\item Protostellar: Clumps visible at 70 $\mu$m with no counterparts at 3 to 8 $				\mu$m within 12 arcsec of the center of the clump, may have counterparts at 24 				$\mu$m and with an unresolved emission at all mid-infrared.
	\item Young Stellar Objects (YSO): Clumps visible at 3-8 $\mu$m, 24 $\mu$m and 					has an unresolved emission at all mid-infrared wavelengths.
	\item H II regions: Clumps visible at 8, 24 and 70 $\mu$m wavelengths, either 				compact or extended with providence of radio emission with the extended 3 to 					8 $\mu$m, infrared bright and associated with high-mass stars.
	\item Photodissociation Regions (PDRs): Areas that display infrared emission but may 		not have pronounced submillimeter continuum, possibly heated externally by 					nearby massive stars or H II regions. 
\end{enumerate}

\noindent As a result, these five distinct clump phases were identified among the ATLASGAL dense massive clumps discussed by \cite{Heyer+etal+2016}. The FWHM linewidths of these clumps were traced through high-density molecules such as $\mathrm{NH_3}$, $\mathrm{HCO}^+$, $\mathrm{N_2H}^+$, HCN, and CS. The classification scheme revealed a total of fifteen quiescent clumps, seventy-seven protostellar clumps, seventy-four YSOs, eight H II regions, and five PDRs. Moving forward, we will conduct in-depth analyses of these 179  clumps to examine the effects of gravity, turbulence, and magnetic fields . The relevant parameters for this study will be estimated in Section 3.

\section{Estimation of Clumps Physical Properties}
\noindent The clump linear size $L_{\mathrm{cl}}$ is calculated as (\citealt{Liu+etal+2022b})
\begin{equation}
	\frac{\sqrt{a.b}}{3600}\left(\frac{\pi}{180}\right)D
\end{equation}
where $a$ and $b$ are the semi-major and minor axes of the clumps (in arcseconds), respectively. $D$ is the kinematic distance of the clumps (in kpc).

Based on the connection between mass and mass surface density $\Sigma_{\mathrm{H_2}}$, we calculated the radius ($R_{\mathrm{cl}}$) of the clump as follows:
\begin{equation}
	R_{\mathrm{cl}} = \sqrt{\frac{M_{\mathrm{cl}}}{\pi\Sigma_{\mathrm{H_2}}}}.
\end{equation}
Eq. (4) was used to estimate the density of clump $\rho_{\mathrm{cl}}$, as defined by Eq. (5).
\begin{equation}
	\rho_{\mathrm{cl}} = \frac{3\,M_{\mathrm{cl}}}{4\pi R_{\mathrm{cl}}^3}.
\end{equation}

\noindent The FWHM,  comprises the non-thermal ($\Delta \upsilon_{\mathrm{NT}}$) and the thermal ($\Delta \upsilon_{\mathrm{T}}$) line widths. We calculated the $\Delta\upsilon_{\mathrm{T}}$ line widths of the clumps using \cite{Wienen+etal+2012} approach:
\begin{equation}
	\Delta\upsilon_{\mathrm{T}} = \sqrt{\frac{8\,\ln (2)\,k\,T_{\mathrm{kin}}}{m_{\mathrm{mol}}}},	
\end{equation}
where $T_{\mathrm{kin}}$ is the kinetic temperature in kelvin (K), $k$ is the Boltzmann constant in joules per kelvin (J/K), and $m_{\mathrm{mol}}$ is the mass of the molecular line tracer in kilograms (kg). For clumps whose $T_{\mathrm{kin}}$ were not available in \cite{Wienen+etal+2012}, we made use of their dust temperature $T_{\mathrm{dust}}$, as found in \cite{Contreras+etal+2013}, to calculate the $\Delta\upsilon_{\mathrm{T}}$. The line width is dominated by $\Delta \upsilon_{\mathrm{NT}}$. The average $\Delta\upsilon_{\mathrm{NT}}$ is $\sim\,2.07\,{\mathrm{km}}\,\mathrm{s}^{-1}$, as compared to $\Delta \upsilon_{\mathrm{T}} \sim\,0.18\,{\mathrm{km}}\,\mathrm{s}^{-1}$. The one-dimensional velocity dispersion ($\sigma$) is then calculated via $\sigma = \Delta\upsilon_{\mathrm{NT}}/2.355$.

The virial parameter ($\alpha_{\mathrm{vir}}$), is often used for accessing the stability of MCs or its fragments.  MCs fall into two classes based on the critical virial parameter ($\alpha_{\mathrm{cr}}$). For instance, if $\alpha_{\mathrm{vir}}$ exceeds $\alpha_{\mathrm{cr}}$, the MC is termed ``subcritical" and such cloud has the potential of swelling, consequently disintegrating and returning to the ISM. On the other hands, ``supercritical" clouds have $\alpha_{\mathrm{vir}}$ less than $\alpha_{\mathrm{cr}}$ and has the potential of collapsing when disturbed as they are gravitationally bound (\citealt{Kauffmann+etal+2013}).

Nevertheless, $\alpha_{\mathrm{cr}}$ can assume several values based on the environmental conditions. It has been reported that in regions with significant magnetic fields, $\alpha_{\mathrm{cr}}$ usually have values less than two (\citealt{Myers+Goodman+1988b}). On the other hand, it has been observed that $\alpha_{\mathrm{cr}}$ can have values greater than two in clouds with insignificant magnetic fields (\citealt{Kauffmann+etal+2013, Ramirez-Galeano+etal+2022}).  For this study, we adopted the \cite{Evans+etal+2021} model of estimation of the $\alpha_{\mathrm{vir}}$. This model did not consider the potential influences of magnetic fields, external pressures, as well as tidal forces, but focused only on the equilibrium between gravity and turbulence, thereby simplifying the understanding of the basic dynamics at play in MCs capable of forming stars.
\begin{equation}
	\alpha_{\mathrm{vir}} = \frac{2\,E_{\mathrm{kin}}}{|E_{\mathrm{g}}|} = 					\frac{1160\,{\sigma}^2 R}{M} = \frac{209\,\Delta\,{\nu}^2 R}{M},
\end{equation}
where $E_{\mathrm{kin}}$ and $E_{\mathrm{g}}$ are the kinetic and potential energies in joules, respectively. $R$, the radius in $\mathrm{pc}$, $M$, the mass in $M_{\odot}$, $\sigma$, the 1D velocity dispersion and $\Delta\upsilon$, the FWHM line width, both in ${\mathrm{km}}\mathrm{s}^{-1}$.

Furthermore, we recalled the approach of \cite{Li+Burkert+2016} in the estimation of the pressures due to gravity ($P_{\mathrm{grav}}$) and turbulence ($P_{\mathrm{turb}}$), as follows: $P_{\mathrm{grav}} = G(L_{\mathrm{cl}}\rho_{\mathrm{cl}})^2/\pi$, $P_{\mathrm{turb}} = (\sigma^2\rho_{\mathrm{cl}})/3$, and defined $\epsilon_{\mathrm{k}}$, the kinetic energy transfer rate per unit volume (\citealt{Luo+etal+2024a}) as: $\epsilon_{\mathrm{k}} = \sigma^3\rho_{\mathrm{cl}}/L_{\mathrm{cl}}$ (\citealt{Kritsuk+etal+2007}).

\subsection{Energy relationships}
\noindent It has been observed that a self-gravitating object in virial equilibrium, has a virial velocity ($\sigma_{\mathrm{vir}}$) relationship with its mass ($M$) and size ($L$) of the form: $\sigma_{\mathrm{vir}} \propto (M/L)^{0.5}$ (\citealt{Luo+etal+2024a}). The energy loss rate ($\epsilon_{\mathrm{vir}}$) of $\sigma_{\mathrm{vir}}$, as described by \cite{Li+Burkert+2016} is
\begin{equation}
	\epsilon_{\mathrm{vir}} \propto M^{3/2}L^{-5/2}.
\end{equation}

Assuming spherical symmetry, with $M \propto \rho L^3$, we obtain
\begin{equation}
	\epsilon_{\mathrm{vir}} \propto \rho^{3/2}L^2.
\end{equation}
\noindent The turbulence energy loss rate in the medium may be represented as (\citealt{Kolmogorov+1941}):
\begin{equation}
	\epsilon_{\mathrm{turb}} \propto \sigma^3L^{-1}.
\end{equation}

If a MC remains in a state of virial equilibrium at all levels of density hierarchy, the rate of transfer of the two forms of energy is equivalent (\citealt{Li+Burkert+2016}):
\begin{equation}
	\epsilon_{\mathrm{vir}} \backsimeq \epsilon_{\mathrm{turb}} .
\end{equation}

\noindent Leading to the corresponding relationship,
\begin{equation}
	\rho^{3/2}L^2 \propto \sigma^3L^{-1}.
\end{equation}

Given that MCs follow the density profile of Eq. (2), the turbulent velocity ($\sigma$), which represents the velocity dispersion observable, can be expressed as follows:
\begin{equation}
	\sigma \propto L^{1-\beta/2}.
\end{equation}

Equations (2) and (13) refer to the comprehensive Larson-type relationships for ($\sigma$-$L$) and ($\rho$-$L$), which are built on two key principles proposed by \cite{Luo+etal+2024a}. First, it’s believed that massive clusters  possess viralized structures that span a wide range of sizes. Second, both virial and turbulent velocities play an equal role in transferring total kinetic energy.

Notably, the Larson relations established in 1981 (see Eqs. 1 and 2) provide precise solutions for these scaling relationships. However, the varying exponents documented in Eq. (1) ($\alpha$ ranging from approximately 0.09 to 0.70, as noted by \citealt{Larson+1981, Fuller+Myers+1992, Plume+etal+1997, Padoan+etal+2016, Traficante+etal+2018b, Zhou+etal+2022, Luo+etal+2024b, Barman+etal+2025}) suggest that alternative formulations, such as those presented by \cite{Luo+etal+2024a}, may also be at play.

This highlights the need to delve deeper into these overall relationships (specifically Eqs. 2 and 13) within actual MCs, especially in regions where high-mass stars are forming. Such research could vastly enhance our understanding of the dominant forces competing in star formation, particularly for those tantalizing high-mass stars.

\subsection{Larson-Type Relationships in the Dense Massive Clumps}
\noindent Here, analyses of the density-size and velocity dispersion-size relationships of our sample will be made and the results discussed.

\begin{figure}[h]
    \centering
    \includegraphics[width=10cm]{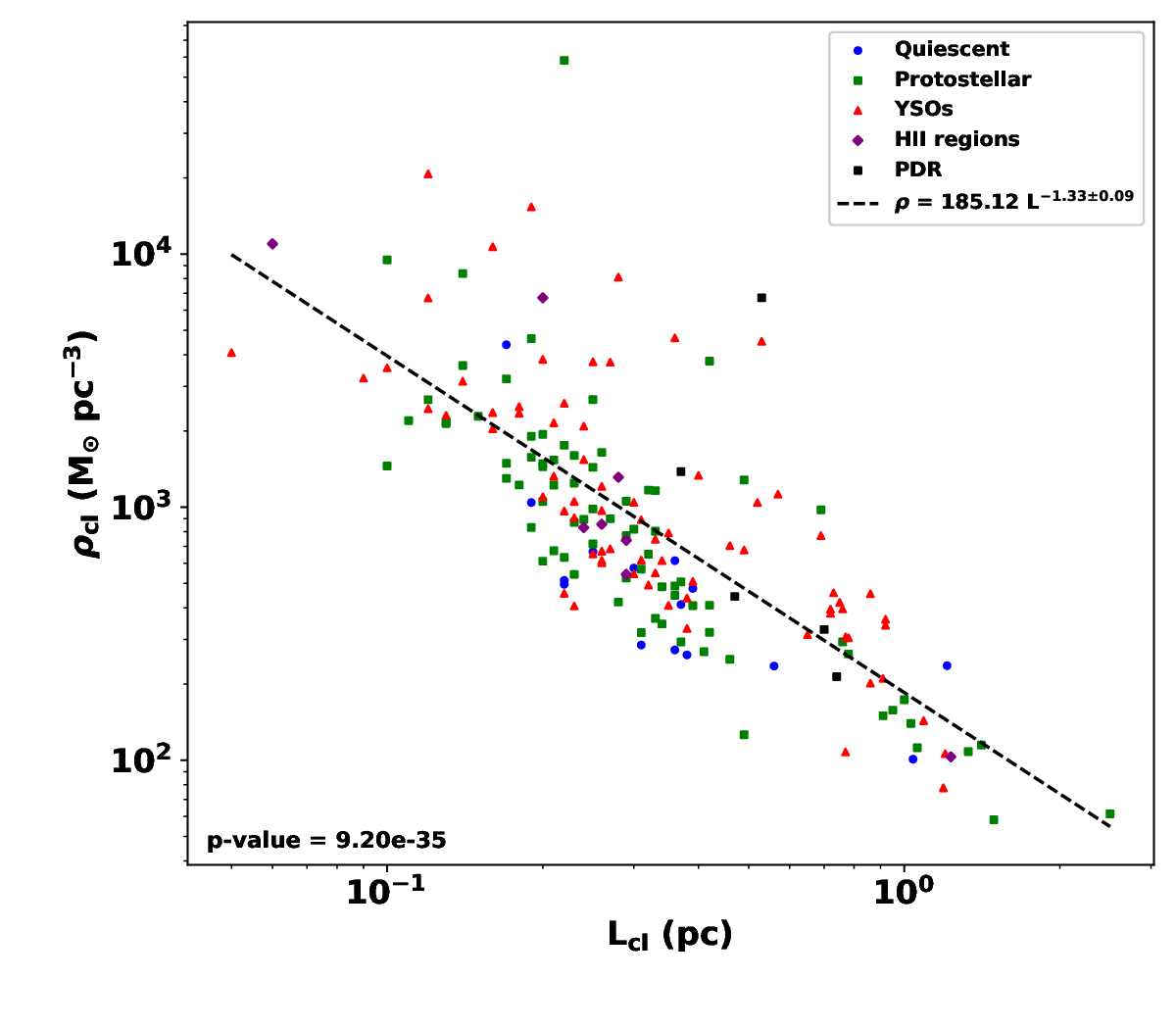}
    \caption{The clumps $\rho_{\mathrm{cl}}$-$L_{\mathrm{cl}}$ relation. Various symbols represent the distinct clumps' evolutionary phases. The dark dashed line denotes the line of best-fit. It is marked with a strong correlation, with Pearson correlation coefficient $r = -0.76$.}
    \label{fig:P-L}
\end{figure}

Fig. 1 illustrates the density-size relation, which represents the density form of the dense massive clumps. The density of these clump structures can be described by the equation  $\rho_{\mathrm{cl}} = \rho_0 (L_{\mathrm{cl}}/\mathrm{pc})^{-\beta}$, where $\rho_0 = (185.12 \pm 36.60)\,M_{\odot}\,\mathrm{pc}^{-3}$ and  $\beta = (1.33 \pm 0.09)$. By substituting the relationship between $\rho_{\mathrm{cl}}$ and $L_{\mathrm{cl}}$, which indicates that $\rho_{\mathrm{cl}}$ is inversely proportional to $L_{\mathrm{cl}}$ raised to the power of $-1.33$ into Eq. (13), we can derive the expected $\sigma$-$L$ relationship for our dense massive clumps:
\begin{equation}
	\sigma \propto L^{0.34}.
\end{equation}

In a recent study, a density-size slope of $-1.54$ was reported for hierarchical MC structures (\citealt{Luo+etal+2024a}). However, \cite{Barman+etal+2025}, observed that $\beta = -1.34$ and $-2.02$, for MCs and cores, respectively. For our clump sample, an exponent $\beta = -1.33$, similar to what was obtained by \cite{Barman+etal+2025} for MCs was observed. These $\beta$ slopes denote a distinct declining trend, indicating that density decreases with an increase in size, in agreement with earlier studies (e.g., \citealt{Larson+1981}; \citealt{Solomon+etal+1987}; \citealt{Heyer+Brunt+2004}; \citealt{MacLow+Klessen+2004}; \citealt{McKee+Ostriker+2007}; \citealt{Lombardi+etal+2010}; \citealt{Mao+etal+2020}; \citealt{Lu+etal+2022}; \citealt{Renaud+etal+2024}).

\cite{Larson+1981} initially observed that giant molecular clouds (GMCs) exhibit comparable column densities. However, it has been contended (e.g., \citealt{Traficante+etal+2018b}) that the density-size relation may only reflect an observational bias stemming from the molecular tracer used in early GMC observations (e.g., \citealt{Kegel+1989}; \citealt{Ballesteros-Paredes+2006}; \citealt{Heyer+etal+2009}). For instance,  \cite{Lombardi+etal+2010} used extinction as a tracer of molecular gas to show that the density-size relation is evident in neighboring MCs just at a certain surface density threshold (\citealt{Traficante+etal+2018b}). The relationship is not applicable to clumps and cores situated inside individual clouds, and an apparent density-size correlation may be considered an artifact of clumps constrained by column density thresholds (\citealt{Ballesteros-Paredes+etal+2012}; \citealt{Camacho+etal+2016}; \citealt{Traficante+etal+2018a, Traficante+etal+2018b}). Numerous studies of large clumps have indicated that they exhibit a variation of about two orders of magnitude in surface densities (\citealt{Urquhart+etal+2014b}; \citealt{Traficante+etal+2015}; \citealt{Svoboda+etal+2016}; \citealt{Elia+etal+2017}; \citealt{Traficante+etal+2018a, Traficante+etal+2018b}).

However, the surface densities of the 179 clumps analyzed in this study (refer to Table 1) range from 192 to 9422 $M_{\odot}\,\mathrm{pc}^{-2}$, exhibiting more than one order of magnitude variation in protostellar and YSOs, while showing less than an order of magnitude variation in quiescent, H II regions, and PDR, likely due to their limited sample sizes. Median values of [443, 568, 678.5, 566, 592] $M_{\odot}\,\mathrm{pc}^{-2}$ were identified in the quiescent, protostellar, YSOs, H II regions, and PDR evolutionary phases, respectively. The limited numbers of certain categories (e.g., quiescent, H II regions, and PDR) prevented the exploration of the relationship between the median values of the surface densities of the clumps. However, a clear increase in median surface density from quiescent to YSOs is obvious.

\begin{figure}[h]
    \centering
    \includegraphics[width=12cm]{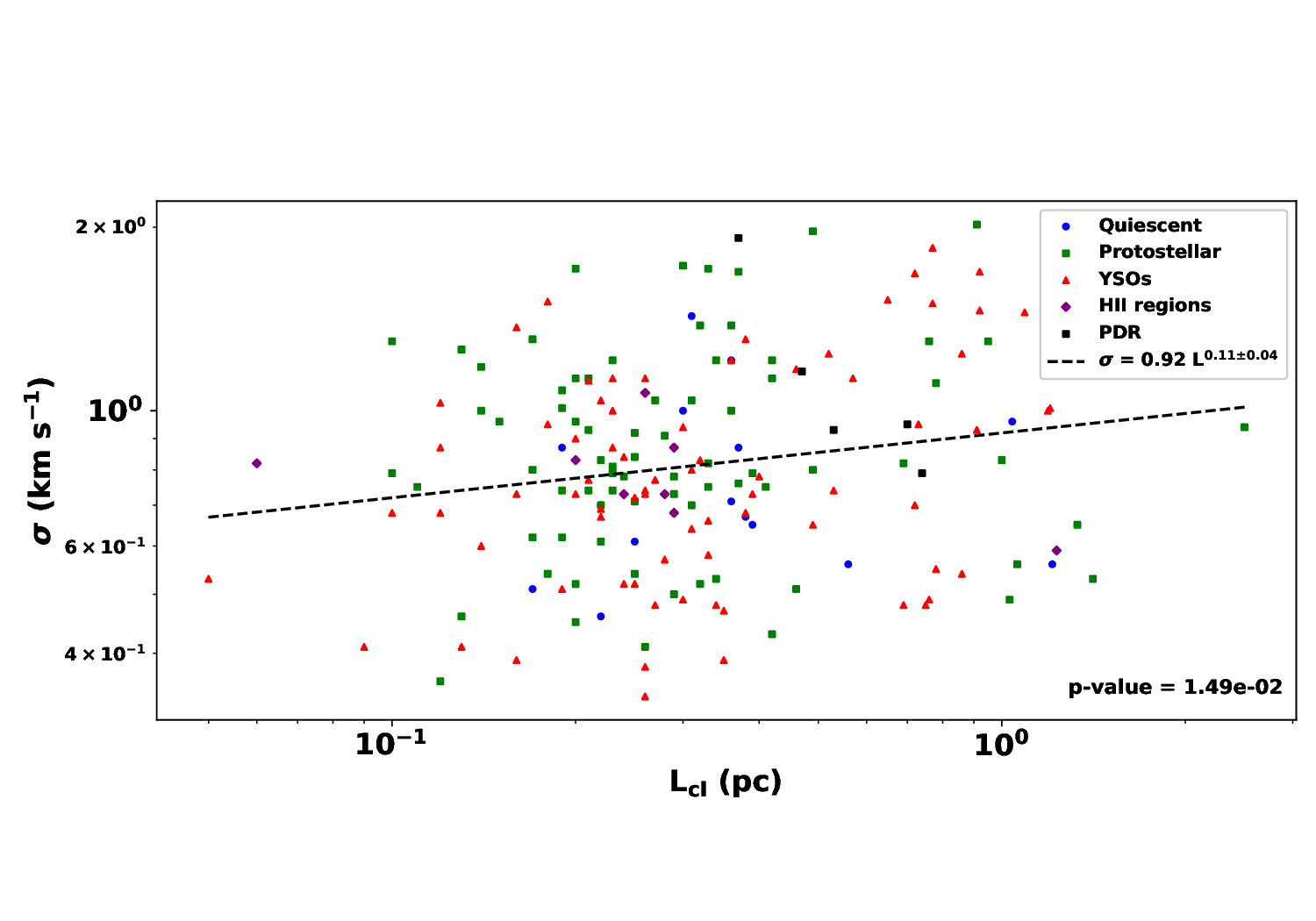}
    \caption{Illustrates the relationship between velocity dispersion and size for the clumps. Various symbols represent the distinct clumps’ evolutionary phases (see Fig. 1). The dark dashed line denotes the best-fit line marked with a weak $r = 0.18$.}
    \label{fig:sigma-L}
\end{figure}

Fig. 2 shows the $\sigma$-$L_{\mathrm{cl}}$ relationship of our sample. This relationship can be described by the equation $\sigma = \sigma_0 (L_{\mathrm{cl}}/{\mathrm{pc}})^{\alpha}$, where $\sigma_0 = (0.92 \pm 0.09)\,{\mathrm{km}}\,{\mathrm{s}}^{-1}$ and $\alpha = (0.11 \pm 0.04)$. The p-value ($\text{1.49e-02}$) shows that the velocity dispersion-size relation is statistically significant, but it is quite weak. The correlation is very weak (with a Pearson correlation coefficient $r = 0.18$) and the slope is almost flat ($\alpha = 0.11$). This suggests that the size of the clump only has a little effect on the velocity dispersion. The weak trend and large scatter suggest that the classical Larson relation does not apply to these dense massive clumps, indicating that forces other than turbulence, like gravity or feedback, are more important in determining their behavior. Nonetheless, \cite{Traficante+etal+2018b} have also reported a comparable observation. The Pearson correlation coefficient ($r$) measures the linear relationship between two variables, ranging from $-1$ to $1$. Here, $r = -1$ shows complete anti-correlation, $r = 1$ shows complete correlation, and $r = 0$ shows no correlation. Notably, the $r \backsimeq 0.18$ shows a weak association between velocity dispersion and size.

Current studies of the velocity dispersion-size relationship in the Galactic MC structures spanning several size scales, have found $\alpha$ exponents ranging from 0.08-0.70 (\citealt{Benedettini+etal+2020, Benedettini+etal+2021}; \citealt{Ma+etal+2021}; \citealt{Lu+etal+2022}; \citealt{Neralwar+etal+2022}; \citealt{Spilker+etal+2022}; \citealt{Zhou+etal+2022}; \citealt{Dong+etal+2023}; \citealt{Li+etal+2023}; \citealt{Feng+etal+2024}; \citealt{Luo+etal+2024a}; \citealt{Barman+etal+2025}; \citealt{Jiang+etal+2025}; \citealt{Yang+etal+2025}). This indicates that the first Larson relation seems to break in some star-forming regions, either with a significantly lower $\alpha$ exponent (e.g., \citealt{Benedettini+etal+2020}; \citealt{Luo+etal+2024a, Luo+etal+2024b}; \citealt{Yang+etal+2025}), or without any correlation especially in denser cores and clumps (e.g., \citealt{CanutGarduno+2024}; \citealt{Barman+etal+2025}). It is to be noted that the different $\alpha$ exponents as reported in the literature, suggests that different object types may exhibit different $\alpha$ values depending on their scale and environment or observational bias. The change in the $\alpha$ exponents between including or excluding GMCs could be due to intrinsic turbulent variation from GMC scale to clump and/or core scale or observational bias since the velocity dispersion measurements may be obtained with different molecular lines and different kind of telescopes (\citealt{Li+etal+2023}).

\cite{Luo+etal+2024b} stated that if the various $\alpha$ exponents which were obviously shallower than the universal first Larson relation were as a result of turbulence in MC structures, the turbulent energy would cascade with a decreasing scale shallower than that reflected from the universal first Larson relation. This could be due to additional inputs of turbulent motions for example by local star-forming feedback (e.g., stellar winds, outflows) especially on smaller scales such as clumps and cores (\citealt{Luo+etal+2024b}). Alternatively, the shallower $\alpha$ exponents could be a natural consequence of the cloud collapse, since it can increase nonthermal motions of the clouds, which are more structured than random turbulence (e.g., \citealt{Ballesteros-Paredes+etal+2018}; \citealt{Traficante+etal+2018b}; \citealt{Ibanez-Mejia+etal+2022}; \citealt{Luo+etal+2024b}).

With our observed shallow $\alpha$ exponent (see Fig. 2), we align with the \cite{Luo+etal+2024b} proposition, of the shallow $\alpha$ being possibly as a result of turbulence and could be due to additional inputs of turbulent motions for instance by local star-forming feedback (e.g., stellar winds, outflows). There is also a possibility that the FWHM linewidths measurements obtained with different molecular lines and different kind of telescopes for our clump sample and the effects of magnetic fields can contribute to the shallow $\alpha$ we observed.

\cite{Traficante+etal+2018b} reported a similar exponent ($\alpha = 0.09$) for massive clumps, noting that the first Larson relation typically does not apply to clump sizes, and this result is not attributed to different internal factors within these clumps. However, our observed exponent ($\alpha = 0.11$) is shallow in contrast to hitherto documented values, such as \cite{Larson+1981} with $\alpha = 0.38$, and \cite{Solomon+etal+1987}, as well as \cite{Heyer+Brunt+2004} with $\alpha = 0.50$. The relationship $\sigma \propto L_{\mathrm{cl}}^{0.11}$ that we measured is much less steep than the expected theoretical scaling of $\sigma \propto L_{\mathrm{cl}}^{0.34}$ (Eq. 14), which is based on the concept of virial equilibrium and layered density structures where $\rho_{\mathrm{cl}} \propto L_{\mathrm{cl}}^{-1.33}$.

This discrepancy suggests that the clumps may not be fully virialized across all spatial scales. The flattening of the slope could indicate a disruption in the balance between turbulent support and gravitational collapse. No doubt, this may be an indicative of the impacts of magnetic or external pressures or due to local dynamical evolution. Additionally, factors such as beam dilution, projection effects, and potential sample bias toward specific clump substructures may contribute to the observed differences. Consequently, turbulence may not be the only factor responsible for the stability of the clumps, as other physical factors may contribute significantly to the behavior and evolution of the clumps.

\begin{figure}[h]
    \centering
    \includegraphics[width=10cm]{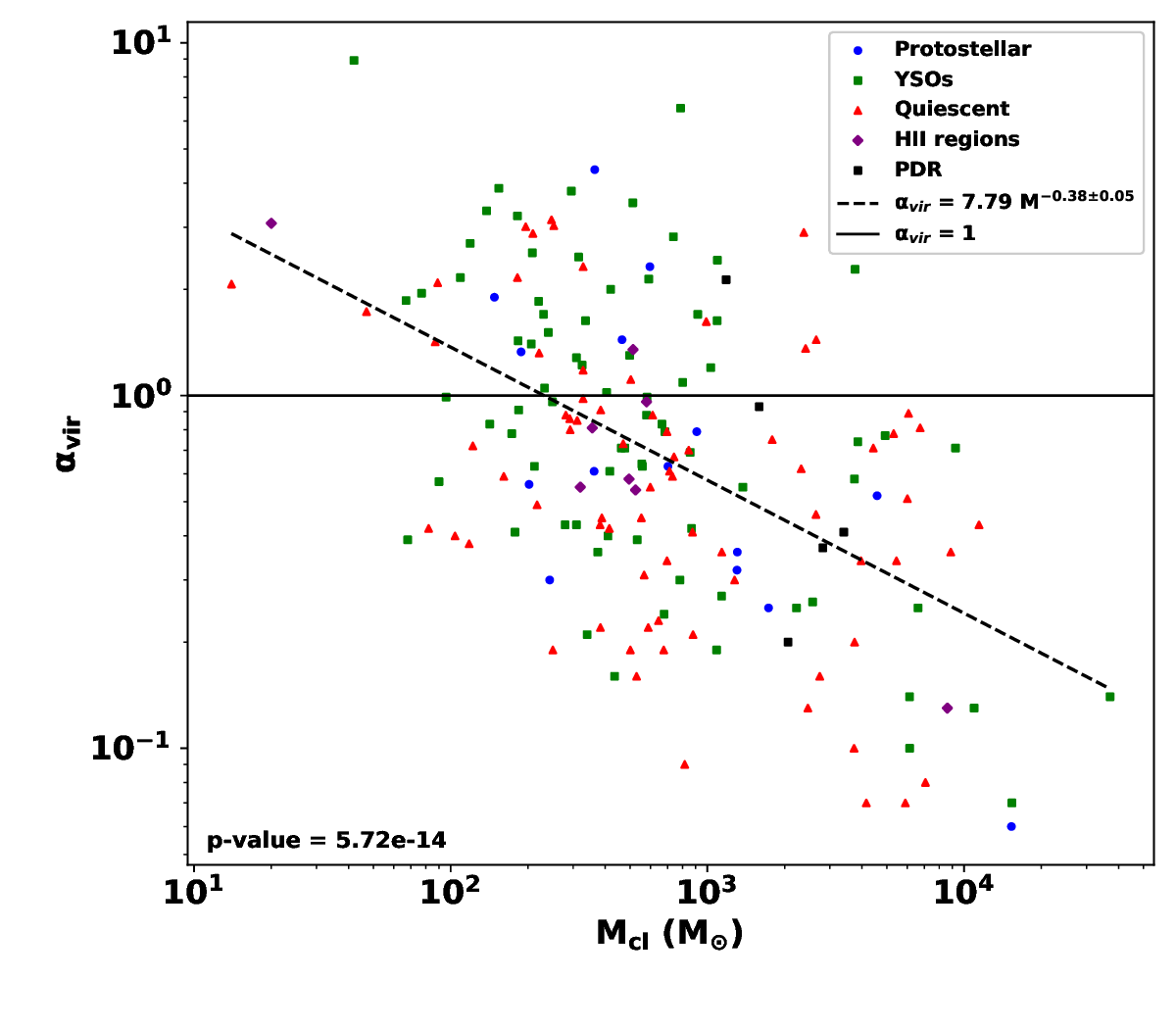}
    \caption{The virial parameter-mass of clumps relationship. The objects are as defined in the first figure. The dark horizontal line denotes unity, while the dark dashed line shows the line of best-fit, with a  moderate $r = -0.52$.}
    \label{fig:vir-M}
\end{figure}

\subsection{The virial parameter-mass of the clumps relationship}
\noindent The comparative significance of self-gravity and kinetic energy has been examined since Larson's study in 1981, which indicated that the virial parameter (refer to Eq. (7)) of MCs often approximates unity (e.g., \citealt{Ballesteros-Paredes+etal+2020}). Numerous studies have used a single molecular line tracers like $^{12}$CO, $^{13}$CO, C$^{18}$O, N$_2$H$^{+}$, NH$_3$, to estimate nonthermal motions and  have revealed a range of virial ratios, with each indicating that $\alpha_{\mathrm{vir}} \propto M^{-\delta}$, where $0 \lesssim \delta \lesssim 1$ (e.g., \citealt{Carr+1987}; \citealt{Loren+1989}; \citealt{Bertoldi+McKee+1992}; \citealt{Urquhart+etal+2014c}; \citealt{Miville-Deschenes+etal+2017}; \citealt{Traficante+etal+2018b}; \citealt{Ma+etal+2021}; \citealt{Lu+etal+2022}; \citealt{Dong+etal+2023}; \citealt{Yang+etal+2025}).

The second Larson relation indicates that GMCs are generally in a state of virial equilibrium. The virial parameter (e.g., \citealt{Traficante+etal+2018b}) is frequently regarded as indicative of the balance between kinetic energy and gravitational energy, under the assumption that other forces, such as magnetic fields, are absent and that the density distribution is spherical and homogeneous (\citealt{Bertoldi+McKee+1992}). Virial equilibrium indicates that $\alpha_{\mathrm{vir}} = \alpha_{\mathrm{eq}} = 1$ (e.g., \citealt{Traficante+etal+2018b}). Alternatively, when modeling a collapsing cloud as an isothermal (Bonnor-Ebert) sphere, hydrostatic equilibrium is approximated at $\alpha_{\mathrm{eq}} \simeq 2$ (\citealt{Kauffmann+etal+2013}; \citealt{Tan+etal+2014}). GMCs are anticipated to achieve virial equilibrium, wherein the kinetic energy from local turbulence counteracts gravitational collapse (\citealt{McKee+Tan+2003}; \citealt{Heyer+etal+2009}). The emergence of massive clumps during a gravo-turbulent collapse (\citealt{Traficante+etal+2018b}) is anticipated to take place within a framework of global virial equilibrium (\citealt{Lee+Hennebelle+2016}).

On the other hand, the observed nonthermal motions may partially stem from the collapse itself and may not inherently counteract gravitational forces (e.g., \citealt{Ballesteros-Paredes+etal+2018}; \citealt{Traficante+etal+2018b}; \citealt{Ibanez-Mejia+etal+2022}; \citealt{Luo+etal+2024b}). In this understanding, virial equilibrium forfeits its original significance (\citealt{Traficante+etal+2018b}). The areas would be in roughly virial equipartition (implying $\alpha_{\mathrm{eq}} = 2$) but misconstrued as being in virial equilibrium (\citealt{Ballesteros-Paredes+2006}).

Regardless of the interpretation of the observed $\alpha_{\mathrm{vir}}$, there is a general agreement that regions with $\alpha_{\mathrm{vir}} < \alpha_{\mathrm{eq}}$ are gravitationally bound and likely to collapse unless supported by strong magnetic fields that can provide stabilization (e.g., \citealt{Kauffmann+etal+2013}; \citealt{Traficante+etal+2018a, Traficante+etal+2018b}). The second Larson relation is not observed in these regions (\citealt{Traficante+etal+2018b}).

The virial parameter of our sample of 179 clumps (refer to Table 1) ranges from $0.06$ to $8.92$. Concerning the clumps' five evolutionary phases, we noted the ranges $0.06 \leq \alpha_{\mathrm{vir}} \leq 4.37$, $0.07 \leq \alpha_{\mathrm{vir}} \leq 8.92$, $0.07 \leq \alpha_{\mathrm{vir}} \leq 3.15$, $0.13 \leq \alpha_{\mathrm{vir}} \leq 3.08$, and $0.20 \leq \alpha_{\mathrm{vir}} \leq 2.13$ for the quiescent, protostellar, YSOs, H II regions, and PDR, respectively. Each evolutionary phase evidently encompasses a minimum of one order of magnitude, with indistinct distinctions among the different phases. A total of 58 clumps possess $\alpha_{\mathrm{vir}} \geq 1$, whereas 29 clumps exhibit $\alpha_{\mathrm{vir}} \geq 2$. The bulk of our clumps are gravitationally bound; nevertheless, if they lack robust magnetic fields (see Sect. 4.1), they fail to achieve gravitational equilibrium. If the kinetic energy arises from turbulence counteracting gravity, its contribution is inadequate to halt or mitigate the collapse at our clump sizes ($0.05$-$2.50$ pc). Therefore, most of our clump samples violate the second Larson relation, which is consistent with the findings of \cite{Traficante+etal+2018b}.

By using our  observed $\sigma$-$L_{\mathrm{cl}}$ and $\rho_{\mathrm{cl}}$-$L_{\mathrm{cl}}$ relationships (specifically, $\rho \propto L^{-1.33}$ and $\sigma \propto L^{0.11}$) in Eq. (7), where $M \propto \rho L^3$ and $R \propto M^{0.5}$, we can find how $\alpha_{\mathrm{vir}}$ relates to mass like this:
\begin{equation}
	\alpha_{\mathrm{vir}} = M^{-0.37}.
\end{equation}

This study estimates the nonthermal motions of 179 clump sample using different tracers (NH$_3$, HCO$^{+}$, N$_2$H$^{+}$, HCN, and CS), resulting in a considerable anti-correlation ($r = -0.52$), as seen in Fig. 3. In Fig. 3, the relationship observed for the clumps can be expressed as $\alpha_{\mathrm{vir}} = \alpha_0 (M_{\mathrm{cl}}/M_{\odot})^{\delta}$, where $\alpha_0$ is ($7.79 \pm 0.83$) and the exponent $\delta$ is ($-0.38 \pm 0.05$). The $\delta$ value is similar to the predicted power in Eq. (15).  This observed inverse relationship between the virial parameter and mass suggests that the more massive a clump, the more it is held by the gravitational force, and this has been confirmed by other studies (e.g., \citealt{Urquhart+etal+2014c}; \citealt{Traficante+etal+2018b}; \citealt{Yang+etal+2025}). This observed pattern shows that the more massive clumps are not in virial equilibrium as their gravitational energy exceeds that of their internal kinetic energy and are dynamically evolving possibly towards global collapse or hitherto in advanced phases of SF. This observed trend challenges the Larson-like theories that clumps are uniform and stable, but instead support the mass dependent clump behavior, likely influenced by factors as gravity, turbulence or magnetic fields. 

It has been contended that the virial parameter-mass anti-correlation is applicable to a collection of clumps, where nonthermal motions are assessed using a singular gas tracer (\citealt{Traficante+etal+2018b}), and appears to dissipate when disparate surveys of clumps and cores, observed with various tracers, are amalgamated (\citealt{Kauffmann+etal+2013}). This prevalent selection skews the observations towards areas with analogous volume densities inside each clump, irrespective of the clump's physical characteristics (i.e., mass and size) (\citealt{Traficante+etal+2018b}).

However, it is important to highlight here that our results shown in Fig. 3 contest previous research (\citealt{Traficante+etal+2018b}), which claims that the virial parameter-mass anti-correlation applies to clumps alone when nonthermal motions are measured using a singular gas tracer.

Furthermore, in stable systems characterized by fast-moving turbulence, the internal ram pressure can be compared with self-gravitational pressure (\citealt{Li+Burkert+2016}). Fig. 4, compares the pressures.

\begin{figure}[h]
    \centering
    \includegraphics[width=10cm]{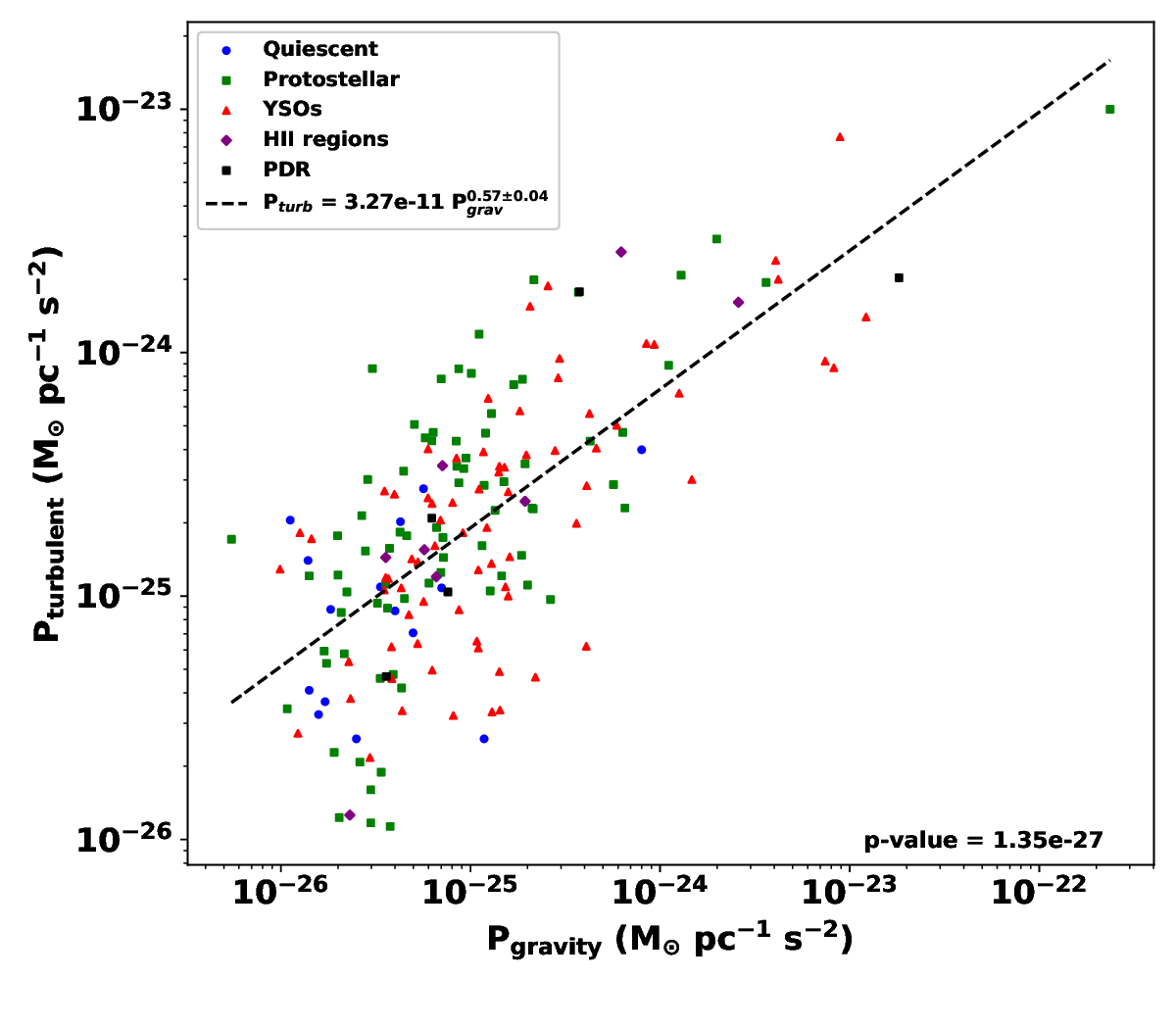}
    \caption{Turbulent pressure as a function of gravitational pressure in the clumps.}
    \label{fig:Pturb-Pgrav}
\end{figure}

\noindent The two pressures illustrated in Fig. 4 can be effectively described by the equation $P_{\mathrm{turb}} = P_0 (P_{\mathrm{grav}}/M_{\odot}\,{\mathrm{pc}}^{-1} {\mathrm{s}}^{-2})^{\gamma}$, with $P_0 = (32.74 \pm 3.30) \times 10^{-12}\,M_{\odot} {\mathrm{pc}}^{-1}\,{\mathrm{s}}^{-2}$, and $\gamma = (0.57 \pm 0.04)$. The strong Pearson correlation coefficient of ($r = 0.70$) indicates a significant relationship, demonstrating that turbulent pressure is proportional to gravitational pressure raised to the power of $0.57$.

\cite{Luo+etal+2024a} found that the value of $\gamma$ is $0.85$ in MC structures, which is slightly higher than the $0.68$ reported for clouds by \cite{Barman+etal+2025}. The $\gamma = 0.68$ indicates that the turbulent ram pressure exerted on the clouds somewhat exceeds the gravitational pressure (\citealt{Barman+etal+2025}). The $\gamma = 0.57$ (see Fig. 4) obtained in our clump sample is less pronounced than the value reported by \cite{Barman+etal+2025} and indicates that turbulence grows with gravity, but insufficiently to counterbalance it, implying that turbulence only partly sustains clumps, whereas gravity assumes more significance for denser and more massive clumps. Such an observation has been reported by \cite{Traficante+etal+2018b}. This pattern indicates that these clumps are likely undergoing global gravitational collapse rather than maintaining equilibrium. Furthermore, magnetic fields oppose gravitational pressure (e.g., \citealt{Barman+etal+2025}), while other factors such as stellar feedback and outflows are intricately linked to the collapse process (e.g., \citealt{Kim+etal+2011}; \citealt{Dobbs+etal+2014}).

In addition, the observed relationship in Fig. 4 supports the schemes of SF, demonstrating how the support due to turbulence declines, emphasizing the several physical factors involved and aligns with a model where gravity dominates turbulence as the clumps evolve. If this dynamic is well understood, can render immense insights into the processes regulating SF and how the interstellar matter works.

\section{The spectrum of turbulent energy}
\noindent Interstellar turbulent gas has often be grouped into two forms:

\begin{enumerate}
	\item Incompressible turbulence, referred to as Kolmogorov turbulence 							(\citealt{Kolmogorov+1941}), is defined by an exponent ($\alpha$) of about 					$\tfrac{1}{3}$, with an energy spectrum ($E(k)$) expressed as $E(k) \propto 				k^{-5/3}$.
	\item Compressible turbulence, referred to as Burgers turbulence 								(\citealt{Li+Burkert+2016}), is defined by an exponent ($\alpha$) of $\tfrac{1}				{2}$ with $E(k) \propto k^{-2}$.
\end{enumerate}
\noindent In these equations, the variable $k$ represents the wave number, which is inversely correlated with $L$. However, the shallow exponent ($\sigma \propto L_{\mathrm{cl}}^{0.11}$) observed in Fig. 2 does not align with the characteristics of either compressible or incompressible turbulence. This discrepancy is evident in the energy spectrum of turbulence as presented (Fig. 5).

\begin{figure}[h]
    \centering
    \includegraphics[width=10cm]{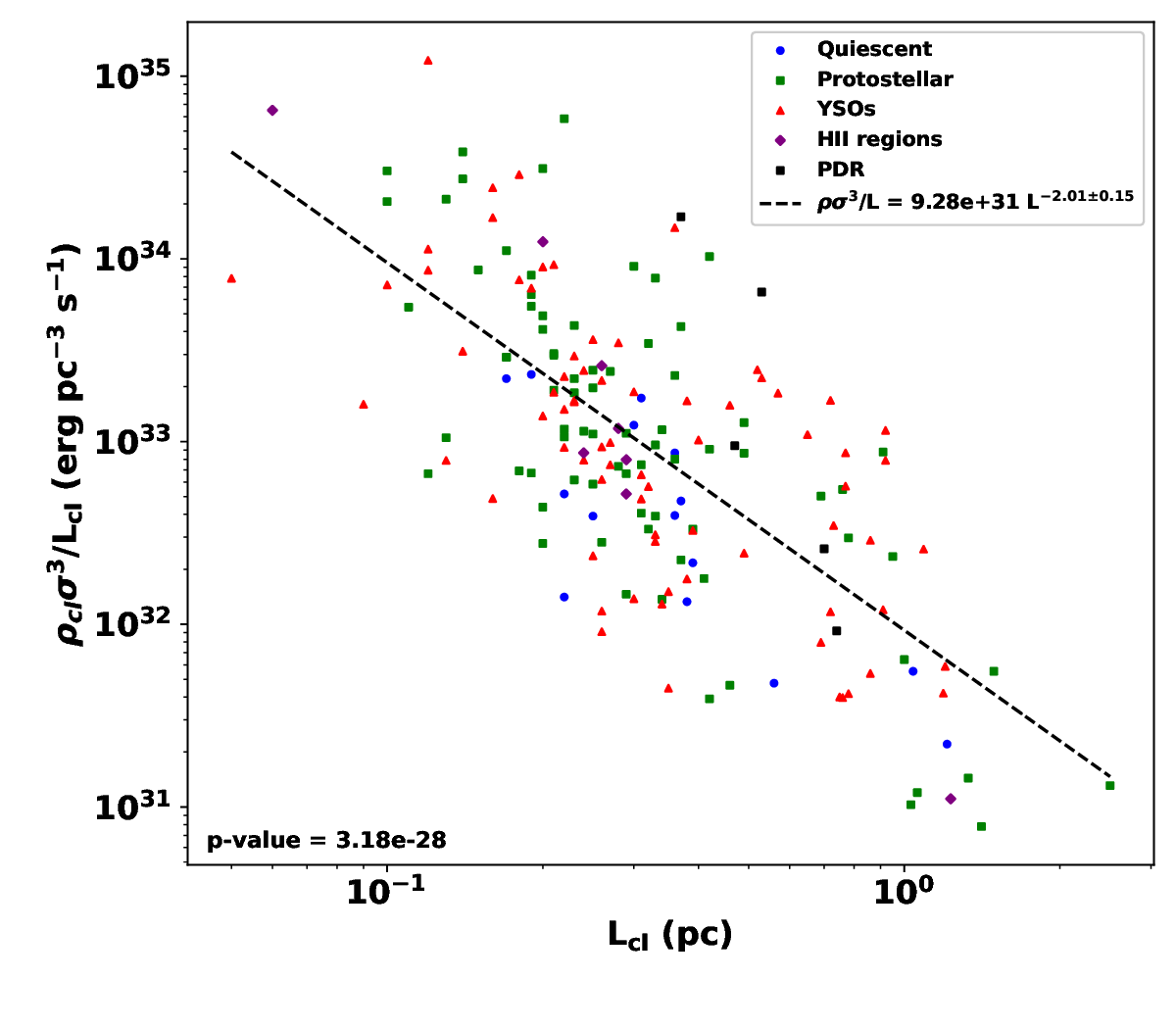}
    \caption{Turbulent kinetic energy transfer rate per unit volume-size relationship of the dense massive clumps. The symbols are the same as shown in Fig. 1.}
    \label{fig:Eturb-L}
\end{figure}

\noindent When kinetic energy begins to flow or disperse, it manifests as turbulence, a dynamic and chaotic phenomenon. ``Exciting simulations of large 3D isothermal supersonic Euler turbulent fluids", have revealed a fascinating insight: that $\epsilon_{\mathrm{k}} = \rho\sigma^3/L$, remains remarkably constant in the ISM (\citealt{Luo+etal+2024a}). This intriguing finding was initially proposed by \cite{Fleck+1988} and has since been reinforced by the research of \cite{Kritsuk+etal+2007} and \cite{Hennebelle+Falgarone+2012}. It illustrates the complex yet fascinating behavior of turbulence in the vast reaches of space.

The observed relationship in Fig. 5 can be described with Eq. (16).

\begin{equation}
	\epsilon_{\mathrm{k}} = \epsilon_0 (L_{\mathrm{cl}}/{\mathrm{pc}})^{-c},
\end{equation}
\noindent in this context, $\epsilon_0 = (9.28 \pm 3.25) \times 10^{31}\,{\mathrm{erg}}\,{\mathrm{pc}}^{-3}\,{\mathrm{s}}^{-1}$ and $c$ = (-2.01 $\pm$ 0.15). However, the combination of Eq. (2) and the fact that $\epsilon_{\mathrm{k}} \propto (L_{\mathrm{cl}}/{\mathrm{pc}})^{-c}$ (see Eq. 16) leads to the expression of the $\sigma$-$L$ relation as
\begin{equation}
	\sigma \propto L^{(1+\beta-c)/3}.
\end{equation}
\noindent Moreover, recalling that $k \propto L^{-1}$, as earlier stated, leads to the expression of Eq. (17) as follows:
\begin{equation}
	\sigma \propto k^{-(1+\beta-c)/3}.
\end{equation}
\noindent Nevertheless, following \cite{Fleck+1988}, the energy spectrum of the turbulence function was expressed as
\begin{equation}
	E(k) = \frac{1}{2}\frac{d\sigma^2}{dk},
\end{equation}

\noindent The use of Eq. (18) in Eq. (19) results in
\begin{equation}
	E(k) = \frac{-(1+\beta-c)}{3} k^{-(5+2\beta-2c)/3}.
\end{equation}
\noindent The substitution of the exponents $\beta = 1.33$ and $c = 2.01$ in Eq. (20) leads to
\begin{equation}
	E(k) \propto k^{-1.21}.
\end{equation}

The measured turbulent energy spectrum $E(k) \propto k^{-1.21}$ in our dense massive clumps has a slope that is less steep than anticipated for both incompressible and compressible turbulence. The traditional \cite{Kolmogorov+1941} theory for incompressible turbulence predicts a spectral index of $\tfrac{-5}{3}$ (meaning $E(k) \propto k^{-1.67}$), while simulations of fast, compressible turbulence show slightly steeper or similar slopes, usually between $-1.8$ and $-2.0$, depending on how the turbulence is driven (like whether it is solenoidal or compressive; see \citealt{Federrath+etal+2010}).

The shallow slope of $-1.21$ indicates that a significant portion of the turbulent kinetic energy exists at large spatial scales, implying a turbulence cascade that is either not fully established or is perpetually replenished at these scales by gravitational contraction, external flows, or stellar feedback.  It has also been reported in the literature that in gravity-dominated systems, turbulence can markedly diverge from classical expectations (\citealt{Kritsuk+etal+2007}; \citealt{Federrath+2013}).

\subsection{Effects of the magnetic fields}
\noindent At this point, we will evaluate the potential impact of the magnetic fields on the stability of our sample.

Based on our aforementioned discoveries, it is feasible some of the clumps, specifically the massive ones, are undergoing collapse by their own gravity. Nevertheless, there is also a possibility that their collapse maybe opposed by significant magnetic support.

Observationally, it has been demonstrated by \cite{Crutcher+2012}, that magnetic support may not adequately counterbalance gravitational support especially in very dense regions (regions with Hydrogen number density ($n_{\mathrm{H_2}}$) greater than $300\, {\mathrm{cm}}^{-3}$), while it may provide a substantial contribution in lower-density areas. \cite{Crutcher+2012} posits an anticipated upper threshold for the strength of magnetic field $B_{\mathrm{up}}$, that has a direct relationship with $n_{\mathrm{H_2}}$, so defined: $B_{\mathrm{up}} \lesssim 150\,\mu G (n_{\mathrm{H_2}}/10^4 {\mathrm{cm}}^{-3})^{0.65}$. \cite{Kauffmann+etal+2013} substituted $n_{\mathrm{H_2}}$ with $\tfrac{3\,M}{4\pi R^3}$, in $B_{\mathrm{up}}$, resulting in
\begin{equation}
	B_{\mathrm{up}} \lesssim 336\,\mu G \left(\frac{M}{10\,M_{\odot}} \right)^{0.65} \left(\frac{R}{0.1\,{\mathrm{pc}}} \right)^{-1.95}.
\end{equation}

\cite{Kauffmann+etal+2013} showed that the non-thermal movements we see are directly related to the strength of the magnetic field needed to keep a clump stable (which is the same as $\alpha_{\mathrm{vir}} = 2$).
\begin{equation}
	B_{M_{\mathrm{BE}}} = 81\,\mu G \left(\frac{M_{\Phi}}{M_{\mathrm{BE}}} \right) \left(\frac{\sigma}{{\mathrm{km\,s^{-1}}}} \right)^2 \left(\frac{{\mathrm{pc}}}{R} \right),
\end{equation}
\noindent where $M_{\Phi}$ and $M_{\mathrm{BE}}$ are the masses of the magnetic flux and sphere in hydrostatic balance (Bonnor-Ebert sphere (BES)), respectively and their ratio (see the first bracket of Eq. 23) equivalently is $\tfrac{2}{\alpha_{\mathrm{vir}}}-1$ (\citealt{Kauffmann+etal+2013}).

\begin{figure}[h]
    \centering
    \includegraphics[width=10cm]{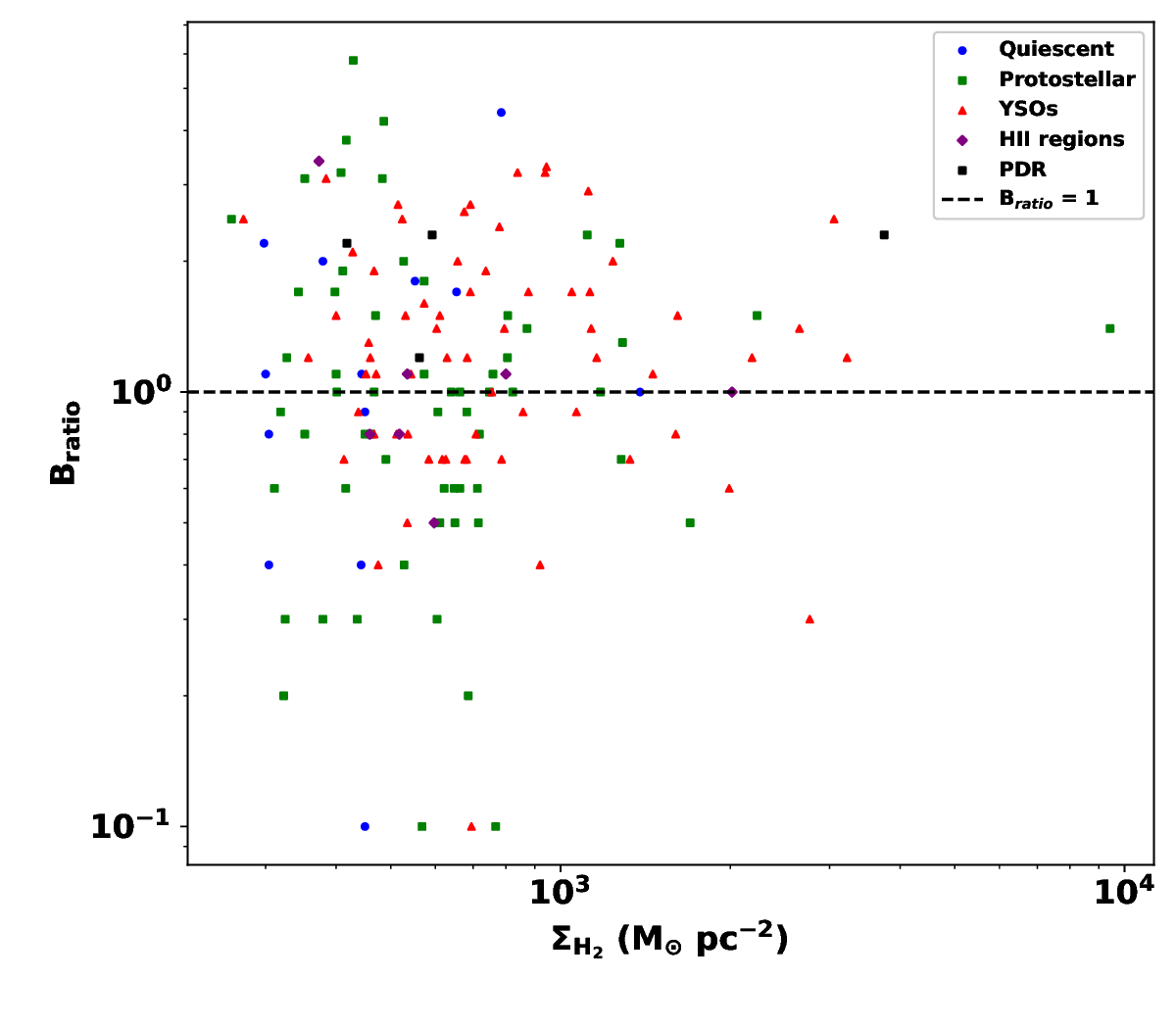}
    \caption{Illustrates the $B_{\mathrm{ratio}}$, defined as the ratio of $B_{M_{\mathrm{BE}}}$ to $B_{\mathrm{up}}$, plotted against surface density. To stabilize with magnetic fields, clumps with a $B_{\mathrm{ratio}}$ over 1 (shown by the dark dashed line) need magnetic field strengths that are higher than the maximum limits suggested by \cite{Crutcher+2012}.}
    \label{fig:Bratio-density}
\end{figure}

Fig. 6 shows how $B_{\mathrm{ratio}} = B_{M_{\mathrm{BE}}}/B_{\mathrm{up}}$ relates to surface density for the 151 clumps with $\alpha \leq 2$. Among these, 46 percent, or 70 clumps, have $B_{\mathrm{ratio}} \leq 1$. On the other hand, most of them, which is 81 clumps or 54 percent, have $B_{\mathrm{ratio}} > 1$, indicating that only very strong magnetic fields can stop them from collapsing. Among these, 46 percent, or 70 clumps, demonstrate a $B_{\mathrm{ratio}} \leq 1$. Conversely, the majority, totaling 81 clumps or 54 percent, present a $B_{\mathrm{ratio}} > 1$, suggesting that only exceptionally strong magnetic fields possess the capacity to prevent their collapse.

\begin{figure}[h]
    \centering
    \includegraphics[width=10cm]{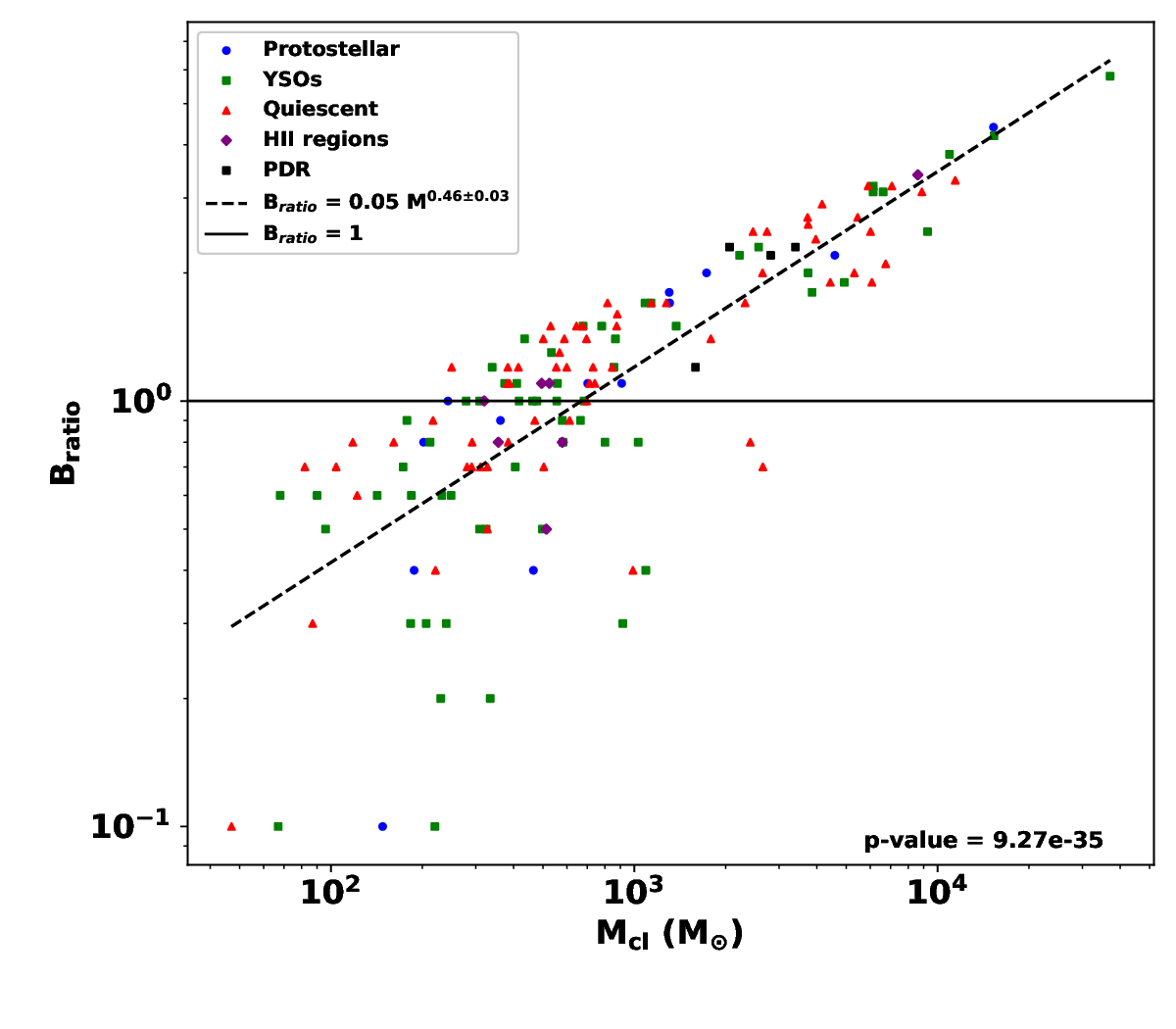}
    \caption{Illustrates the relationship between $B_{\mathrm{ratio}}$ and mass. The dark dashed line denotes the best-fit line with a strong correlation ($r = 0.80$), meaning that as the mass of the clumps increases, the magnetic fields needed to keep them stable also increase significantly. In the case of $M_{\mathrm{cl}} \geq 3000\,M_{\odot}$, it is imperative that extraordinarily strong magnetic fields are present to impede the process of collapse.}
    \label{fig:Bratio-mass}
\end{figure}

Fig. 7 shows the relationship between $B_{\mathrm{ratio}}$ and the mass of the clumps. The figure shows that the strength of the magnetic fields needed to keep a collapsing clump stable is higher than what \cite{Crutcher+2012} predicted when the clump has a mass of $3000\,M_{\odot}$ or more.

The findings indicate that magnetic fields may play a role in certain clumps; however, for most, especially the more massive ones, this mechanism alone is insufficient to maintain collapse at the scale of the clumps. This study recognizes that, in the inner cores, magnetic fields can have a strong effect and help prevent gravitational collapse (\citealt{Fontani+etal+2016}).

\subsection{Caveats}
\noindent Temperature is a critical factor influencing the chemical and physical properties of dense cores in interstellar clouds and their potential progression toward star formation. The chemistry and dust properties exhibit temperature dependence; thus, interpreting any observation necessitates an understanding of temperature and its fluctuations. Molecular line spectroscopy allows us to directly measure the gas kinetic temperature, and $\mathrm{NH_3}$ is the most commonly used marker for this. As mentioned in Section 3, for clumps that don't have $T_{\mathrm{kin}}$ data from \cite{Wienen+etal+2012}, we used the dust temperature ($T_{\mathrm{dust}}$) from \cite{Contreras+etal+2013} to calculate the $\Delta\upsilon_{\mathrm{T}}$. This calculation should influence their $\sigma$ and $\alpha_{\mathrm{vir}}$. Nonetheless, this issue is addressed as $T_{\mathrm{kin}}$ and $T_{\mathrm{dust}}$ can thermally couple at high densities, $n({\mathrm{H_2}}) \backsim 10^5 {\mathrm{cm}}^{-3}$ and higher (see \citealt{Goldsmith+2001}).

Additionally, our discussions about how size relates to velocity dispersion and density might be affected by observational biases because our data comes from various studies using different telescopes and molecular tracers. By consistently calculating the needed parameters (like size, radius, velocity dispersion, and virial parameter) as outlined in Sect. 3, we might lessen this effect somewhat. Considering these potential biases, we will focus on the size of clumps in relation to their specific stage of evolution. By consistently calculating the necessary parameters (e.g., size, radius, velocity dispersion, virial parameter), as specified in Sect. 3, this impact might be reduced in some way. Considering these possible biases, we will focus on the size of clumps in relation to their specific stage of evolution. The same study can't be done as accurately as it could be because there is a limited range of clump sizes and more observational biases affecting the local scaling analysis. We also advocate that to have better comparisons of the velocity dispersion, density, and size relations in MCs, together with their corresponding turbulent energy spectrum (\citealt{Luo+etal+2024a}), future research should make use of the same telescope as well as molecular tracer to consistently observe MCs kinematics at several scales.

Therefore, this study is limited by the fact that the clump properties are derived from various surveys and molecular tracers that investigate distinct density regimes. The velocity dispersion-size, density-size, and turbulence spectrum may indicate survey-dependent systematics instead of a consistent physical trend. The characteristics of turbulence should be interpreted cautiously, as they do not originate from a homogeneous observational dataset (see, e.g., \citealt{Forgan+Bonnell+2018}).

Also, detailed observations using tools like ALMA or NOEMA are important for a complete study of the magnetic field characteristics of these clumps at core sizes.

\subsection{Inference on the formation of high-mass stars}
\noindent When studying clumps that have the likelihood of the formation of high-mass stars ($M_{\star} > 8\,M_{\odot}$), the mass-radius diagram is an invaluable tool (\citealt{Traficante+etal+2018b}). Fig. 8 displays the mass-size diagram of our clumps. 

\begin{figure}[h]
    \centering
    \includegraphics[width=10cm]{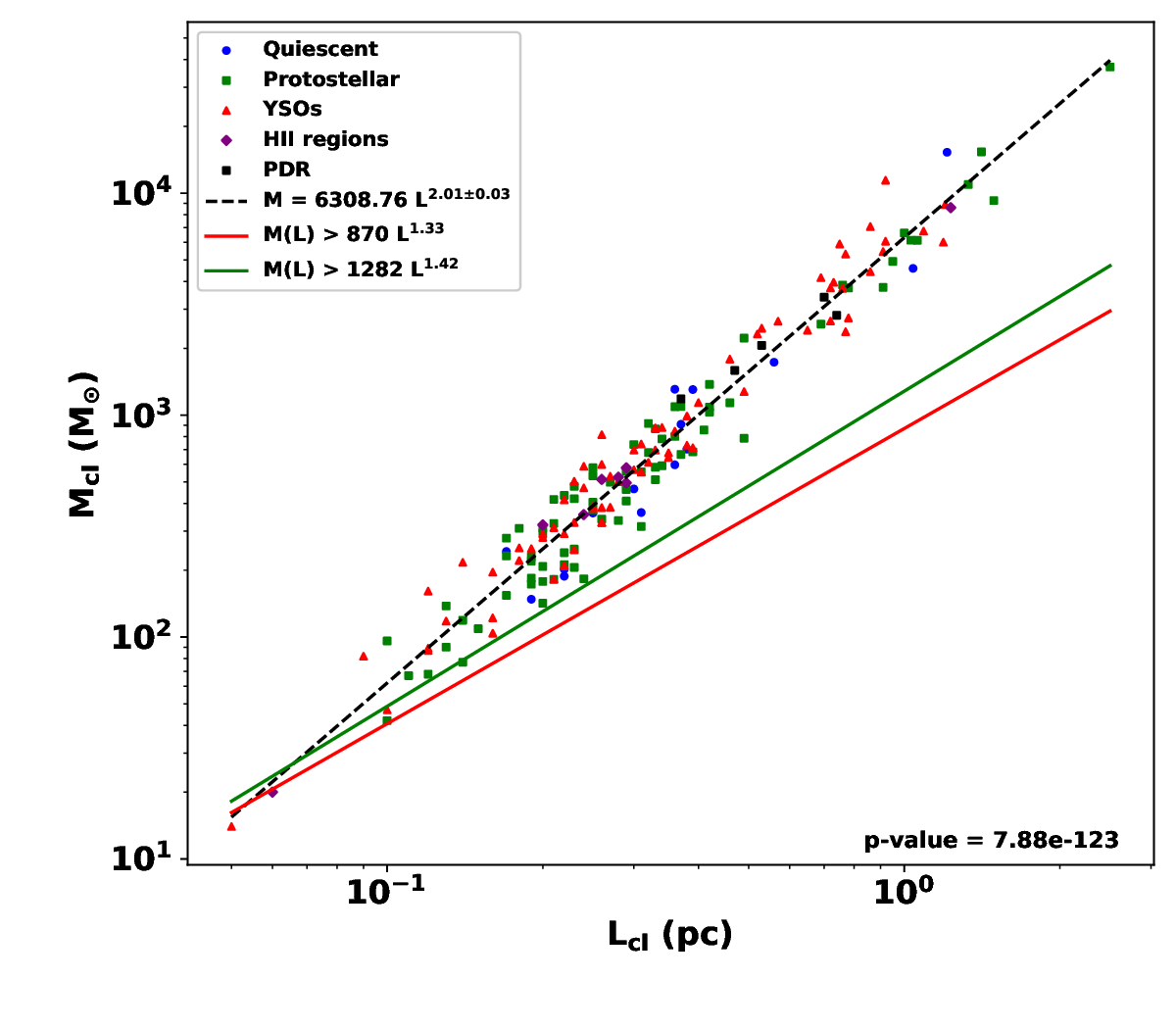}
    \caption{Mass-size distribution of the clumps. The high-mass SF limit, $M(L) > 870\,M_{\odot}\,(L/{\mathrm{pc}})^{1.33}$, of \cite{Kauffmann+Pillai+2010} is shown by the red line, whereas the updated \cite{Baldeschi+etal+2017} limit, $M(L) > 1282\,M_{\odot}\,(L/{\mathrm{pc}})^{1.42}$, is displayed by the green line. The dark dashed line is the best-fit line for our sample, $M_{\mathrm{cl}} \propto L_{\mathrm{cl}}^{(2.01 \pm 0.03)}$, with a very strong correlation $r = 0.98$.}
    \label{fig:Mass-Size}
\end{figure}

Fig. 8 illustrates that the predominant number of these clumps is capable of forming high-mass objects, in accordance with the empirical mass-size criteria as established (\citealt{Kauffmann+Pillai+2010, Baldeschi+etal+2017}). Only but two of the clumps (AGAL030.368+00.484 and AGAL346.369-00.647), all are over the former threshold, and except for five clumps (AGAL012.988+00.354, AGAL022.559+00.171, AGAL030.368+00.484, AGAL040.907-00.249, and AGAL346.369-00.647), all are above the latter, which represents a more rigorous criterion.

Fig. 8 can also be used to obtain an alternative perspective on the third Larson relation (\citealt{Traficante+etal+2018b}). A sample of star-forming regions exhibiting approximately constant column density is expected to have a mass distribution characterized by $M \propto R^{\eta}$ where $\eta$ is approximately equal to 2 (\citealt{Traficante+etal+2018b}). Previous surveys of massive clumps have identified a wide range of values for $\eta$, which is significantly influenced by the various methodologies employed to derive the dust properties of the clumps (\citealt{Traficante+etal+2018b}). Mass-radius diagrams exhibit slopes ranging from $ \eta \simeq 1.6$ to $1.9$ (\citealt{Lombardi+etal+2010}; \citealt{Kauffmann+etal+2010}; \citealt{Urquhart+etal+2014c}; \citealt{Barman+etal+2025}; \citealt{Yang+etal+2025}), as well as slopes of $\eta \geq 2$ (\citealt{Ellsworth-Bowers+etal+2015}; \citealt{Traficante+etal+2018b}; \citealt{Benedettini+etal+2021}; \citealt{Ma+etal+2021}; \citealt{Lu+etal+2022}; \citealt{Dong+etal+2023}) and even higher values ($\eta \geq 2.7$; \citealt{Ragan+etal+2009}), with some instances showing no clear correlation, particularly in dense cores (\citealt{Barman+etal+2025}).

The power-law slope of $2.01 \pm 0.03$ from Fig. 8, alongside the slope $\beta = -1.33 \pm 0.09$ from Fig. 1, supports the notion of an approximately constant surface density for our clump sample. This indicates that our clump sample adheres to the third Larson relation, consistent with recent findings for MCs with sizes less than 1 pc (\citealt{Barman+etal+2025}).

However, a notable divergence from the standard Larson-type turbulence scaling is shown by the low correlation ($r = 0.18$) between velocity dispersion and size (see Fig. 2) in our sample of dense massive clumps. This points to the possibility that local mechanisms like feedback, magnetic fields, or gravity affect turbulence at  our clump scales. This feedback mechanism reintegrates material into the ISM and maintains turbulence (e.g., \citealt{Barman+etal+2025}).

Several other mechanisms (\citealt{Barman+etal+2025}), such as accretion, spiral shocks, galactic shear, magnetorotational instability, and gravity-driven turbulence, can also drive ISM turbulence (\citealt{Nakamura+Li+2007}; \citealt{Goldbaum+etal+2011}; \citealt{Padoan+etal+2016}; \citealt{Peters+etal+2017}; \citealt{Elmegreen+Elmegreen+2019}; \citealt{Xie+Li+2025}). Within galaxy disks, this mechanism affects the rate of SF, energy dissipation, and angular momentum transmission (\citealt{Barman+etal+2025}). Swing-amplified instabilities, spiral-arm torques, and in-plane accretion all contribute to gravity-driven SF (\citealt{Barman+etal+2025}) by using galactic potential energy to create the required turbulence (\citealt{Elmegreen+2024}). However, it is important to state here that these other mechanisms are outside the scope of this study.

Multiscale models of high-mass SF may help explain the poor relationship between velocity dispersion and size in our clumps. Within the framework of the GHC model, such dispersion lends credence to the idea that turbulence is not the principal controller of clump dynamics (e.g., \citealt{Ballesteros-Paredes+etal+2011, Vazquez-Semadeni+etal+2019}). Instead of a disordered cascade which is dependent on the size, gravitational collapse dominates the interior motions and the local infall events are represented by the velocity dispersion. This perspective aligns with the time-dependent hierarchical fragmentation which is self-gravity driven. However, the weak correlation in Fig. 2 can be explained by the anisotropic accretion and external converging flows in the I2 model (see, \citealt{Padoan+etal+2020}). In the I2 model, the infusion of momentum into clumps is influenced by unpredictable inflow-driven turbulence, which is generally independent of the clump size. The GHC and I2 models concur that the interior dynamics of dense massive clumps are governed by gravity and accretion flows, rather than conventional turbulence.

\section{Summary and Conclusions}
\noindent This study examined the effects of gravity, turbulence, and magnetic fields on 179 clumps across various evolutionary phases. The clumps consisted of fifteen quiescents, seventy-seven protostellar objects, seventy-four young stellar objects (YSOs), eight H II regions, and five photodissociation regions (PDRs). We utilised the Larson-like relations concerning velocity dispersion, density, and size, along with the associated energy spectrum of turbulence and estimates of magnetic fields within the clumps to achieve this objective. We computed and incorporated the p-values in our plots by examining correlations to demonstrate the statistical significance of the relationships analyzed. All p-values indicate that the correlations examined were statistically significant.

A significant deviation from the conventional Larson-type turbulence scaling was observed regarding the relationship between velocity dispersion and size. This suggests that local mechanisms such as magnetic fields, gravity, or feedback may influence turbulence at clump scales. The observed dispersion in the $\sigma$-$L_{\mathrm{cl}}$ relation suggests a significant influence of gravity on the dynamics of high-mass star-forming clumps, reflecting the complex interplay of these factors.

We found the relationships $\sigma$-$L_{\mathrm{cl}}$ and $\rho_{\mathrm{cl}}$-$L_{\mathrm{cl}}$, where $\sigma$ is proportional to $L_{\mathrm{cl}}$ raised to the power of $0.11$, and $\rho_{\mathrm{cl}}$ is proportional to $L_{\mathrm{cl}}$ raised to the power of $-1.33$. The observed relationship $\sigma \propto L_{\mathrm{cl}}^{0.11}$ is notably flatter than the theoretical prediction of $\sigma \propto L_{\mathrm{cl}}^{0.34}$, which is anticipated based on the assumptions of virial equilibrium and a hierarchical density structure denoted by $\rho_{\mathrm{cl}} \propto L_{\mathrm{cl}}^{-1.33}$.  This discrepancy indicates that the clumps of different sizes are not fully virialized. The flattening of the slope may indicate a disturbance in the equilibrium between turbulence support and gravitational force, potentially influenced by external pressure, magnetic fields, or localized changes.

The virial parameter-mass relation we found ($\alpha_{\mathrm{vir}} \propto M_{\mathrm{cl}}^{-0.38}$) closely matches the theoretical prediction ($\alpha_{\mathrm{vir}} \propto M_{\mathrm{cl}}^{-0.37}$). The trend of $\alpha_{\mathrm{vir}} \propto M_{\mathrm{cl}}^{-0.37}$ indicates that more massive clumps exhibit a greater degree of gravitational binding. This inverse relationship signifies a systematic deviation from virial equilibrium, particularly in the high-mass regime, where gravitational potential prevails over internal kinetic support. This trend suggests that large clumps are undergoing dynamic evolution, likely experiencing global collapse or being in advanced stages of SF.

The observed $P_{\mathrm{turb}} \propto P_{\mathrm{grav}}^{0.57}$ indicates that, particularly at higher pressures where gravitational forces dominate, turbulence provides only limited support against self-gravity. This tendency indicates that numerous clumps may transition from virial equilibrium to collapse. The observed modest correlation aligns with scenarios of star formation governed by the gradual depletion of turbulent support, indicating a dynamic range of physical variables and reinforcing a model where gravity ultimately surpasses turbulence during clump evolution.

Our turbulent energy spectrum study, using the velocity dispersion-size along with the density-size relationships, shows that $E(k)$ is proportional to $k^{-1.21}$, which means a lot of the turbulent kinetic energy is found at larger scales. This suggests that the turbulence cascade is either not completely formed or is constantly being renewed at these scales by gravitational contraction, outside flows, or feedback from stars.

Our findings showed that, on average, magnetic fields do not play a significant role in the stability of the clumps. Remarkably substantial magnetic fields is required in the stabilization of clumps having masses greater than or equal to $3000\,M_{\odot}$.

We equally showed that the predominant number of these clumps is capable of forming high-mass objects, in accordance with the established empirical mass-size criteria.

\section*{Expressions of Gratitude}
\noindent We express our gratitude to the anonymous referees for their remarks and recommendations that significantly enhanced the quality of this research. This research has received assistance from the Pan-African Planetary and Space Science Network (PAPSSN), an Academic Mobility initiative (grant No. 624224) financed by the Intra-African Mobility Scheme of the European Education and Culture Executive Agency (EACEA). PAPSSN seeks to establish a mobility program for students, academic personnel, and support staff among partners from Botswana, Ethiopia, Nigeria, South Africa, and Zambia, focusing on the thematic areas of Science, Technology, Engineering, and Mathematics (STEM) and Information and Communications Technology (ICT), with a specific emphasis on Planetary and Space Sciences (PSS).

\bibliographystyle{aasjournal}
\bibliography{bibtex}

\end{document}